\documentclass[amsmath,amssymb,aps,prd,11pt,tightenlines,superscriptaddress, 
nofootinbib,preprintnumbers,notitlepage]{revtex4-1}
 \usepackage{ulem}
\usepackage{graphicx}  
\usepackage{dcolumn}   
\usepackage{bm}        
\usepackage{amssymb}   
\usepackage{amsmath}
\usepackage{mathrsfs}  
\usepackage{graphicx,color} 
    
\usepackage[english]{babel}     
\usepackage{graphicx,wrapfig,lipsum}      
\usepackage{multimedia}    
\usepackage[mathscr]{eucal}  
\usepackage{bm}              
\usepackage{amssymb}        
\usepackage{amsmath}     
\usepackage{mathrsfs}  
\usepackage[mathscr]{eucal}      
\usepackage{xcolor}
\usepackage{sidecap}  
\usepackage{wrapfig}    
\usepackage{subfig}     
\usepackage{float}  

\begin{document}     

\title{The infrared Yang-Mills wave functional due to percolating center vortices}   
 
    
\author{D. R. Junior}
\affiliation{Instituto de F\'\i sica, Universidade Federal Fluminense,   
24210-346 Niter\'oi - RJ, Brasil.}
\affiliation{Institut f\"ur Theoretische Physik, Universit\"at T\"ubingen, 72076 Tübingen, Deutschland.}
\affiliation{INFES, Universidade Federal Fluminense, 28470-000 Santo Ant\^onio de P\'adua - RJ, Brasil.}
\author{L.~E.~Oxman}
\affiliation{Instituto de F\'\i sica, Universidade Federal Fluminense,   
24210-346 Niter\'oi - RJ, Brasil.} 
\affiliation{Institut f\"ur Theoretische Physik, Universit\"at T\"ubingen, 72076 Tübingen, Deutschland.}
\author{H. Reinhardt }  
 \affiliation{Institut f\"ur Theoretische Physik, Universit\"at T\"ubingen, 72076 Tübingen, Deutschland.}

\date{\today}

\begin{abstract} 

    Inspired by the center-vortex dominance in the infrared sector of $SU(N)$ Yang-Mills theory observed on the lattice, we propose a vacuum wave functional localized on an ensemble of correlated center vortices endowed with stiffness and magnetic monopoles that change the orientation of the vortex flux. In the electric-field representation, this wave functional becomes an effective partition function for N complex scalar fields. The inclusion of both oriented and non-oriented vortices as well as so-called N-vortex matchings leads to an effective potential that has only a center symmetry left. In the center-vortex condensed phase, this symmetry is spontaneously broken. In this case, the Wilson loop average can be approximated by a solitonic saddle-point localized around the minimal surface. The asymptotic string tension thus obtained displays Casimir scaling. 



\end{abstract}  
 
\maketitle     

\section{Introduction}

Understanding confinement of quarks and gluons is one of the fundamental problems of particle physics.  Due to its nonperturbative character,  lattice simulations defined in an Euclidean spacetime have been essential to assess this phenomenon in a reliable way. The quest has been focused on the characterization of relevant configurations. In Ref. \cite{indirectcenter}, center vortices were detected as the infrared dominant configurations of Yang-Mills theory. In particular, in Ref. \cite{ELRT} it was shown that the center vortices found after center projection in the maximal center gauge represent indeed physical degrees of freedom, in the sense that  their density shows the proper scaling towards the continuum limit. Furthermore, in the center projected gauge theory, the deconfinement phase transition emerges as a depercolation transition from a phase of percolating vortices to a phase of small vortices predominantly aligned along the time-axis \cite{langfeld_T}. Although the field-strength of center vortices is along the Cartan algebra, they carry topological Pontryagin charge \cite{Reinhardt_topology}, \cite{reinhardtvortices}. A nonzero total topological charge of a center vortex requires the vortex flux to be non-oriented, with the change of orientation generated by magnetic monopole loops on the center-vortex surfaces \cite{Reinhardt_topology}.

In $4$d $SU(N)$ Yang-Mills (YM) theory, the ensemble of percolating center-vortex worldsurfaces detected in center-projected Monte Carlo configurations reproduce an area law with $N$-ality for the Wilson loop \cite{nality}. This type of vortex ensemble has  been modeled in terms of random closed worldsurfaces represented on the lattice by a set of plaquettes \cite{randomsu2, randomsu3}. They are governed by a lattice action with a  term proportional to its area (tension) and another one proportional to the number of pairs of nonparallel neighboring plaquettes (stiffness). In this manner, the confining string tension for fundamental quarks and the order of the deconfinement phase transition were described \cite{randomsu3}. 
In Ref. \cite{diff1} (see also \cite{universe7080253}), it was argued that ensembles of percolating oriented and non-oriented center-vortex surfaces together with natural correlations could  generate, besides $N$-ality, the confining flux tube between quarks \cite{Cosmai-2017,Yanagihara2019210,Kitazawa,su3} and the L\"uscher term \cite{L_scher_2002}. The line of reasoning is as follows. In the lattice formulation of a condensate of loops, which generate closed worldsurfaces in 4d, the Goldstone modes are $U(1)$ gauge link-variables governed by the Wilson action \cite{Rey}. Now,   besides loops, center vortices may form closed arrays where $N$ lines are matched at a given point. In the 4d lattice, this corresponds to configurations where the plaquettes form open worldsurfaces glued at their borders to form closed arrays. This is done  with the condition that plaquettes at the borders are attached in groups of $N$ to a common link. By promoting the gauge link-variables from $U(1)$ to $SU(N)$, this matching rule was taken into account. The inclusion of arrays where center-vortex worldsurfaces are  attached to monopole worldlines, with their own natural matching rules, 
 was done by including an ensemble of adjoint lattice holonomies. Finally, the naive continuum limit led to effective $SU(N)$ gauge fields and minimally-coupled interacting adjoint scalar fields. This is the correct field-content to drive a Spontaneous Symmetry Breaking (SSB) phase that supports a topologically stable flux tube betwen quarks, with $N$-ality, corrected by the collective transverse fluctuations.  Moreover, the Abelian-like profiles observed in YM lattice simulations \cite{Cosmai-2017,Yanagihara2019210,Kitazawa,su3} and the asymptotic Casimir scaling of the string tension, which is among the possible behaviors \cite{teper, 4dlaw}, can also be accommodated in these models \cite{O-V, O-S, O-S-J}. 
  
Establishing the relevance of $N$-matching and correlations with lower dimensional defects is of primary interest to complete the picture of confinement provided by random worldsurfaces. One possible line would be a careful exploration of the path discussed above. For example, adding stiffness is expected to be essential for a well-defined continuum limit of the lattice model in Ref. \cite{Rey} and its possible extensions \cite{diff1}. Otherwise, the surfaces would collapse, as occurs with triangulated random surfaces when only the Polyakov (or Nambu-Goto) action is considered \cite{ambjorn,wheater}.

The aim of this work is to assume center vortex dominance in the Yang-Mills vacuum and to study the effect of various vortex features on the confinement properties measured by the Wilson loop. For this aim, we shall use the Hamiltonian approach (based on the canonical quantization in the Weyl gauge $A_0=0$) where, at a given time, a quantum state is represented by a wave functional $\Psi(A)$ for the spatial components $A_i(x)$, $i=1, 2 , 3$, defined on the physical space $x\in \mathbb{R}^3$. Besides being more transparent, the Hamiltonian approach has the technical advantage that we have to deal only with the simpler one-dimensional loops instead of the two-dimensional vortex surfaces of the Euclidean functional integral approach. Generic center-vortex surfaces in four dimensional space-time emerge in three dimensional space as loops (which, as time passes, trace out the two-dimensional vortex worldsurfaces). We shall assume a vacuum wave functional which is concentrated on an ensemble of correlated elementary center vortices endowed with stiffness, the center-vortex $N$-matching rule, and attached monopoles. The inclusion of stiffness is absolutely necessary. As is well known from the study of random polymers, the end-to-end probability is ill-defined when the monomer size $a$ goes to zero. This can be circumvented by invoking an effective monomer size $a_{\rm eff}$, which incorporates stiffness as the alignment of microscopic  monomers on a finite physical scale \cite{kleinert}. Interestingly, there is also the option of explicitly including stiffness and implementing the continuum limit in the presence of external fields \cite{fred, diff3}. These studies were essential for applications to interacting  ensembles formed by center-vortex worldlines  and monopole worldlines in 3d and 4d Euclidean spacetime, respectively \cite{diff2, O-S-D-3d, oxmanreinhardt}, \cite{diff1}.  They will also prove useful in the  Hamiltonian description of the center vortex ensemble in $3+1$ dimensions, where probability amplitudes for the vortex loops  will be characterized by  properties, like tension and stiffness, inherited from the four dimensional worldsurfaces. Our starting point will be a gas of elementary center-vortex loops. Then, on  top of this, we will include the effect of center-vortex matching 
and attached monopoles.

The organization of the paper is as follows: in section \ref{anatomy} we discuss the representation of  center vortices in terms of Abelian variables. In section \ref{wave funct}, we present a vacuum wave functional peaked at these configurations, including loops as well as arrays formed by correlated center vortices. Section \ref{sectionwilson} is devoted to compute the Wilson loop in this state. Finally, in section \ref{conclusion} we present our conclusions.

 \section{Anatomy of center-vortices } \label{anatomy}
 In 4d Euclidean spacetime, center vortices are field configurations representing closed surfaces of electric or magnetic flux whose Wilson loop $ W[A](C)$
is given by a non-trivial center element provided the loop $C$ is non-trivially linked to the center-vortex surface. In 3d, center-vortex configurations $A(C)$ have flux localized on closed loops $C$. They satisfy
\begin{equation}
    W\left[A(C_1)\right](C_2)=Z^{L(C_1,C_2)}  \;,
    \label{wl2}
\end{equation}
 where $L(C_1,C_2)$ is the Gaussian linking number between the center-vortex loop $C_1$ and the external quark loop $C_2$. The center element $Z$ depends on the quark representation and on the $Z(N)$ vortex charge. General antisymmetric quark representations  (also known as fundamental representations), which are labelled  by the $N$-ality $k$, will be discussed in the Appendix. In the body of this work, we shall consider quarks in the defining representation of $SU(N)$ where  the Wilson loop is
\begin{equation}
    W[A](C)= \frac{1}{N} {\rm Tr} \left( {\rm P}\, \exp{\left[i\oint_{C} dx \cdot A \right]} \right)  \;,
\label{wl1}
\end{equation}
 with the components of the vector field $A$ being  $N\times N$ matrices in $\mathfrak{su}(N)$. 
 Now, even in this representation, the center element $Z$ in Eq. \eqref{wl2} could still assume
$N-1$ different values $Z_l= e^{i\frac{2\pi l}{N}}$,  $l=1,2,...,N-1$, associated with the possible center-vortex charges. The ensembles we shall consider will always involve elementary center vortices, which are characterized by $Z= e^{\pm i\frac{2\pi}{N}}$, whose powers generate the whole $Z(N)$ group. In addition, as changing $C_1 \to -C_1$ changes the sign of the linking number, we can consider, say, 
$ Z= e^{- i\frac{2\pi}{N}}$. In this case, 
\begin{equation}
    Z \, I =e^{i \mathscr{C}}\makebox[.5in]{,} \mathscr{C}=2\pi2N \omega \makebox[.5in]{,} \omega = \omega^q T_q \;,
    \label{z1}
\end{equation}
where  $T_q$, $q=1, \dots, N-1$ are the Cartan generators, while the tuple $\vec{\omega} = (\omega^1,\dots,\omega^{N-1})$ is a weight of the defining representation.
We shall use the term  ``co-weight'' for the algebra-valued quantity $\mathscr{C}$. In fact, in the defining representation of $SU(N)$ there are $N$ different weights, and thus $N$ different co-weights $\mathscr{C}_{[j]}, j=1,2,...,N$, that give rise to one and the same center element $Z= e^{-i\frac{2\pi}{N}}$.  They satisfy the relation
\begin{equation} 
    \sum_{j=1}^N \mathscr{C}_{[j]}=0 \;.
    \label{w2}
\end{equation} 
Throughout the paper we adopt the normalization  $(T_A,T_B)= \delta_{AB}$, where the internal product between two Lie Algebra elements $X, Y$ is given by the Killing form, $( X,Y )= {\rm Tr}({\rm Ad}(X) \, {\rm Ad}(Y))$, where ${\rm Ad} (\cdot)$ denotes the adjoint representation of $\mathfrak{su}(N)$. For the generators in the defining  representation this implies ${\rm Tr}\, (T_AT_B)=\delta_{AB}/(2N)$.

\subsection{Abelian projection}
\label{Abep}

On the lattice, center vortices were studied in the direct maximal center gauge, which brings the link variables $U_\mu \in SU(N)$ as close as possible to center elements $  Z_\mu \in Z(N)$, which were used to define a center-projected lattice \cite{directcenter}. This way, center vortices can be detected as those objects that pierce the P-plaquettes. These degrees of freedom were also studied in the indirect maximal center gauge, where the maximal Abelian gauge is initially used to bring the link variables as close as possible to Cartan (diagonal) variables $ C_\mu $. Next, center vortices can be detected by using the remaining $U(1)^{N-1}$ gauge symmetry to obtain a center-projected lattice out of the Abelian-projected variables $C_\mu$ \cite{indirectcenter}.
In the next section, we shall construct a wavefunctional $  \Psi(A)$ peaked on an ensemble of Abelian projected center vortices, which is aimed at describing the infrared properties of the vacuum state in $SU(N)$ Yang-Mills theory in the Schr\"odinger representation.  In continuum $(3+1)$d spacetime, $\Psi(A)$ depends on the gauge field variable $A(x)$ defined on the real space ($x \in \mathbb{R}^3$).

In order to represent the different center-vortex configurations in this Abelian context, it is convenient to start with an oriented vortex line $\gamma$, associated with a co-weight $\mathscr{C}$, which can be considered as part of a closed center vortex loop. Such a vortex line gives rise to a gauge potential
\begin{align}
   &a_\mathscr{C}(x,\gamma)=-\mathscr{C} \int_{\gamma} d\bar{x} \times \nabla_{x}D(x-\bar{x}) \;,
    \label{thinvortex}
\end{align}
where $D(x)$ is the Green's function of the Laplacian in three dimensions, i.e. $-\Delta D(x) = \delta^{(3)}(x)$\;. This gauge field may also be written in terms of a source $j(x,\gamma)$ localized on the path $\gamma$, as follows
\begin{align}
    a_\mathscr{C}(x,\gamma)=\mathscr{C} \frac{\nabla\times j(x,\gamma)}{-\Delta}\makebox[.5in]{,}j(x,\gamma)=\int_{\gamma} d\bar{x}\; \delta(x-\bar{x})\;.
    \label{a2}
\end{align}
From this representation follows for the magnetic field
\begin{align}
     &\nabla\times a_\mathscr{C}(x,\gamma)=\mathscr{C}  \big(j(x,\gamma) +\nabla (-\Delta)^{-1} \nabla\cdot  j(x,\gamma)\big) \;.
     \label{a3}
\end{align}
For a closed oriented path $\gamma,\; \partial\gamma=0$ we have $\nabla \cdot j(x,\gamma)=0$ so that 
\begin{align}
     &\nabla\times a_\mathscr{C}(x,\gamma)=\mathscr{C} j(x,\gamma)\;,
     \label{a4}
\end{align}
while for an open oriented path $\gamma$ starting (ending) at $x_{\rm i}$ ($x_{\rm f}$) one finds
\begin{align}
     &\nabla\times a_\mathscr{C}(x,\gamma)=\mathscr{C} j(x,\gamma) -\mathscr{C}\frac{x-x_{\rm i}}{4\pi|x-x_{\rm i}|^3}+\mathscr{C}\frac{x-x_{\rm f}}{4\pi|x-x_{\rm f}|^3} \;.
     \label{a5}
\end{align}
 Of course, in the latter case, $a_\mathscr{C}(x,\gamma)$ does not represent a true center vortex, as there is no concept of linking between an open line and a quark loop $C$. Accordingly, the  last two terms in Eq. \eqref{a5} give the contributions from the endpoints, which are the magnetic fields of a magnetic monopole and antimonopole, respectively, with magnetic "charges" $\pm \mathscr{C}$. By conservation of magnetic flux, an open magnetic flux line has to carry a magnetic monopole and antimonopole, respectively, at its endpoints. Then, using Stokes' theorem, the contribution to the exponent of $ W[A](C)$ contains the flux 
generated by the monopoles through a surface $S(C)$ with boundary $C$ plus the intersection number between $\gamma$ and $S(C)$. On the other hand, when $\gamma $ is closed, $a_\mathscr{C}(x,\gamma)$ does correspond to a center vortex. That is, the Wilson loop is given by Eq. \eqref{wl2}, as it only depends on the intersection number, which can be equated to the
linking number $L(\gamma , C)$ in this case. 

\begin{figure}
    \centering
    \includegraphics[scale=0.4]{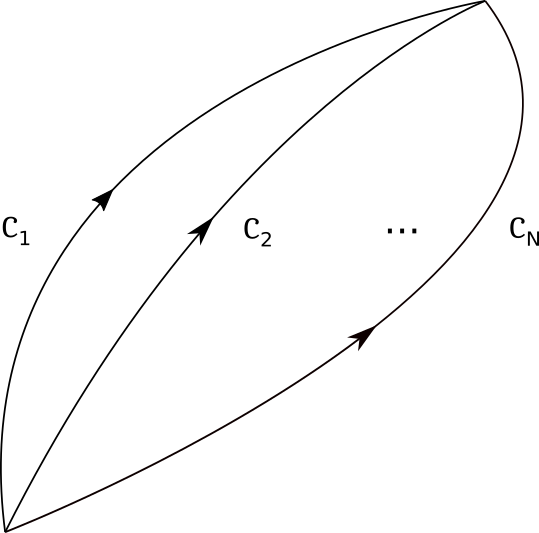}
    \caption{Here, we depict the flux for an array with a pair of $N$-matching points. The arrows indicate the orientation of the paths $\gamma_1,\dots,\gamma_N$, which are associated to the co-weights  $\mathscr{C}_{1},\dots,\mathscr{C}_{N}$, respectively ($\mathscr{C}_{1} + \mathscr{C}_{2} + \dots + \mathscr{C}_{N}=0$).}
    \label{nvortex}
\end{figure}

A closed $\gamma$ can also  be obtained by gluing together two open oriented lines $\gamma_1$, $\gamma_2$ associated with the same co-weight $\mathscr{C}$, such that the monopole endpoint of $\gamma_1$ coincides with the anti-monopole endpoint of $\gamma_2$ and vice versa. In this manner,  the monopole and anti-monopole contributions to the total magnetic field cancel and we are back to Eq. \eqref{a4}. Now, since each co-weight of the defining representation yields by Eq. \eqref{z1} the same center element we can also form center vortices by gluing together open vortex lines carrying different co-weights,
 \begin{align}
a(\mathscr{V}) = \sum_{n} a_{\mathscr{C}_n} (\gamma_n) \;.
\label{a6}
\end{align}
Of particular interest will be vortex configurations $\mathscr{V}$ formed by identifying the monopole or anti-monopole endpoints of $N$ open vortex lines $\gamma_1, \gamma_2 , \dots ,$, each belonging to a different co-weight $\mathscr{C}_n\in\{\mathscr{C}_{[k]}\}$. Due to the property \eqref{w2} of the co-weights the contributions of all $N$ magnetic (anti-)monopoles at such matching points (referred to in the following as $N$-matchings) cancel and  we find for the associated magnetic field
 \begin{align}
\nabla \times  a(\mathscr{V}) = \sum_{n}\mathscr{C}_n\, j(\gamma_n) \;.
\end{align} 
In general, since each open vortex line has two endpoints and $N$ open lines meet at each $N$-matching point a vortex field configuration consisting only of N-point matchings (i.e. no  oriented vortex loops) has to satisfy the sum rule: $2I=NV$ where $I$ is the number of open vortex lines and $V$ is the number of vertices. Here again the Wilson loop only depends on an intersection number that can be equated to a linking number between the closed array and $C$. 
For example, for a pair of $N$-matching-points (see Fig. \ref{nvortex}), the vortex field configuration is given by
\begin{align}
 a(\mathscr{V}) =  \sum\limits_{n=1}^{N}a_{\mathscr{C}_n}(\gamma_n)\;,
    \label{a7}
\end{align}
where $\gamma_1,\dots,\gamma_N$ are open lines all starting and ending, respectively, at the same point (but each being associated with a different co-weight). In this case,  Eq. \eqref{wl2} is obtained with $C_1$ being the composition of $N-1$ loops,
\begin{equation}
C_1 = (\gamma_1- \gamma_N) \cup \dots \cup (\gamma_{N-1} - \gamma_N) \;. 
\end{equation}
There is still another relevant type of matching rule that we will consider. The indirect maximal center gauge allows to analyze not only the center projected link-variables $Z_\mu$ but also the Abelian-projected ones $C_\mu$. While $Z_\mu$ shows the presence of percolating center vortices localized on closed surfaces (resp. loops) in 4d (resp. 3d), the analysis of $C_\mu$ makes it possible to keep track of the orientation of the flux in the Cartan subalgebra. In 4d, for $SU(2)$, besides surfaces characterized by a single orientation,  it was noticed  that the Lie Algebra orientation can change at monopole worldlines \cite{chains}. 
 This was done by applying the De-Grand and Toussaint method, introduced  in Ref. \cite{degrand}  to analyze the link-variable $e^{i\theta_\mu}$ in the compact $U(1)$ gauge theory. In this case, the flux of  $\theta_\mu$ on  a plaquette was written as $f_{\mu \nu} = \bar{f}_{\mu \nu} + 2\pi n_{\mu \nu}$, where $ \bar{f}_{\mu \nu}  \in [-\pi, +\pi]$ and 
$n_{\mu \nu} \in \mathbb{Z}$. In 4d, at a fixed time-slice, as the total flux through the surface of a cube is zero, the total flux of $\bar{f}_{\mu \nu}$ is nontrivial if and only if the total flux of the Dirac field $ 2\pi n_{\mu \nu}$ is nontrivial, which occurs when there is a monopole inside the cube. The plaquettes with $n_{\mu \nu} \neq 0$ can be changed by a gauge transformation, but the monopole locations cannot. These locations are the only physical degrees of freedom associated with $n_{\mu \nu}$. 
In the case of $SU(2)$, the fluxes live in the Cartan subalgebra generated by $\sigma_3$. In general, 
on a cube around  a monopole, $\bar{f}_{\mu \nu}$ could be spread. However, the simulations showed that it is collimated.  There is a flux $\pi\sigma_3$ entering one of its plaquettes and a flux $-\pi\sigma_3$ leaving another. 
The total flux is conserved due to a flux  $2\pi\sigma_3$ carried by a Dirac string leaving a third plaquette. The latter contributes trivially to the Wilson loop and action: $e^{i2\pi\sigma_3 }=I$. Indeed, it was established  that about 61\% of the vortex lines have no monopoles on them, 31\% contain a monopole-antimonopole pair, and 8\% of closed vortex lines have an even number of pairs, with monopoles alternating with antimonopoles \cite{chains}.  For general $N$, the non-oriented case would correspond to situations where a (collimated)
flux $\mathscr{C}_1$ enters a plaquette of a cube around a monopole and a flux $\mathscr{C}_2$ leaves through a different plaquette.\footnote{Note that for $N=2$ the co-weights are $ \mathscr{C}_{[1]} = +\pi \sigma_3$, $\mathscr{C}_{[2]} = -\pi \sigma_3 $. For $N=3$, in terms of the diagonal Gell-Mann matrices, we have $\mathscr{C}_{[1]} = \pi \left(\lambda_3+ \lambda_8/\sqrt{3} \right)  $, $\mathscr{C}_{[2]} = \pi \left(-\lambda_3+ \lambda_8/\sqrt{3} \right)  $, $\mathscr{C}_{[3]} = -2\pi \lambda_8/\sqrt{3}  $. }  The flux would be conserved due to the presence of a nontrivial flux $\mathscr{C}_1 -\mathscr{C}_2$ leaving a third (trivial) plaquette: $e^{i (\mathscr{C}_1 -\mathscr{C}_2)} = Z \bar{Z} \, I = I $ (cf. Eq. \eqref{z1}). Therefore,  it is clear that, in the continuum, if  non-oriented collimated fluxes were described in terms of Cartan  vector gauge fields, the introduction of Dirac strings would be required. In the present work, the use of Dirac strings will be avoided by extending the field content to include a Cartan scalar monopole field.
In this regard, let us
initially consider arrays of lines where the monopole endpoint of one of them coincides with the anti-monopole endpoint of the other. 
For example, we can  take a gauge field $a(\mathscr{V})$ of the form given in Eq. \eqref{a6} constructed in terms of $M$ oriented lines $\gamma_n$, $n=1, \dots M $, carrying weights $\mathscr{C}_n$, which form a chain. They are glued such that the 
final endpoint ($x_{n}^{\rm f}$) of $\gamma_n$ coincides with the initial endpoint of $\gamma_{n+1}$ ($n < M$) and 
the final endpoint $x_M^{\rm f}$ of $\gamma_M$ coincides with the initial endpoint of $\gamma_1$.
\begin{figure}[h]        
\centering 
\subfloat[M=2, {\rm non-collimated}]{ \includegraphics[scale=.4]{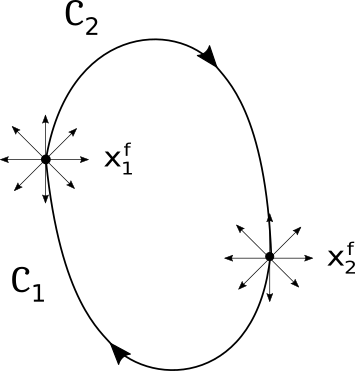}  } 
\hspace{0.9cm}
\subfloat[M=2]{ \includegraphics[scale=.4]{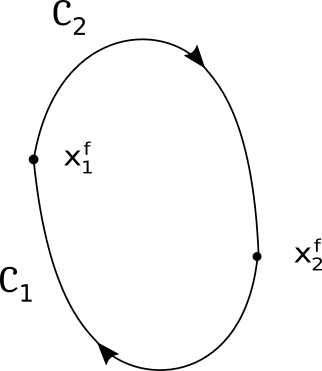}  } \hspace{0.9cm}
	 \subfloat[M=3]{\includegraphics[scale=.4]{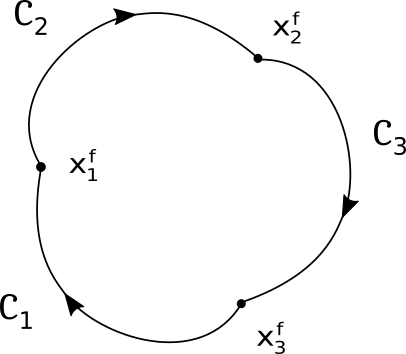}} \hspace{0.2cm}   
\caption{\small In
(a), we show the flux 
$\nabla \times a(\mathscr{V})$ (cf. Eq. \eqref{m2}) for $M=2$. The arrows on the paths indicate their orientation. There is a non-collimated component with flux $\mathscr{C}_1 -\mathscr{C}_2$ ($\mathscr{C}_2 -\mathscr{C}_1$) leaving $x_1^{\rm f}$ 
($x_2^{\rm f}$). In (b), the flux is collimated 
by including Dirac strings $\delta_1$, $\delta_2$  (not displayed) leaving $x_1^{\rm f}$, $x_2^{\rm f}$, and carrying the same fluxes as before. Their unobservability can be implemented by means of appropriate scalar potentials when defining the total flux. Another possible non-oriented (in the Lie algebra) collimated flux, is shown in (c).}    
\label{CV-link}     
\end{figure} 
Thus, following this sequence of lines, a closed path is obtained. In addition, from one line to the next, the flux orientation given by one of the possible co-weights $\mathscr{C}_{[j]}$ changes to a different value. In this case, the associated magnetic flux is not collimated; instead, it is given by
\begin{align}
     &\nabla\times a(\mathscr{V}) = \sum_{n=1}^M\mathscr{C}_n\, j(\gamma_n)  + \sum_{n=1}^M (\mathscr{C}_n-\mathscr{C}_{n+1})\, \frac{x-x_n^{\rm f}}{4\pi|x-x_n^{\rm f}|^3} \;,
     \label{m2}
\end{align}
where $\mathscr{C}_{M+1} = \mathscr{C}_{1}$ (for $M=2$, see Fig. \ref{CV-link}a).
In addition, as the different co-weights give rise to the same center element, the contribution to the Wilson loop originated from the first term in Eq. \eqref{m2} can be equated to the linking number between the closed path $\gamma_1 \cup \dots \cup \gamma_M $ and $C$, which is the fingerprint of center vortices. The second term leads to additional 
solid-angle contributions subtended from the monopole locations. This is not in line with the lattice simulations, as the contribution of non-oriented lattice configurations to the Wilson loop is  given by (the lattice version of) 
Eq. \eqref{wl2}, i.e. by the linking number between the chains and the quark loop. To get Abelian non-oriented collimated configurations, we still have to introduce in $a(\mathscr{V})$ an additional term
\begin{align}
\sum_n a_{\mathscr{E}_n}(\delta_n)  \makebox[.5in]{,} a_{\mathscr{E}_n}(\delta_n) = \mathscr{E}_n   
\int_{\delta_n} d\bar{x} \times \nabla_{x}D(x-\bar{x}) \makebox[.5in]{,} \mathscr{E}_n= \mathscr{C}_n - \mathscr{C}_{n+1}  \;,
    \label{ds}
\end{align}
for Dirac lines $\delta_n$ running from  $x_n^{\rm f}$ to $\infty$. The effect is to  replace Eq.  \eqref{m2} by the  total flux
\begin{align}
 \sum_{n=1}^M  \mathscr{C}_n\, j(\gamma_n) +  \sum_{n=1}^M  \mathscr{E}_n \, j(\delta_n)   \;.
\end{align}
Again, like in the lattice, the Dirac lines do not contribute to the Wilson loop, which only receives the center-vortex contribution originated from the first term. Of course, the treatment that must be given to center-vortex and Dirac lines in the ensemble is completely different. For example,  unlike the former, the latter do not have physical properties such as stiffness and tension. In order to get rid of the unobservable Dirac lines, leaving only the physical effect of their endpoints,  
we shall extend the field content. More precisely, in section \ref{ele-mon}
we shall include a Cartan scalar monopole potential, defining the magnetic flux such that only the physical collimated part $ \sum_{n=1} \mathscr{C}_n\, j(\gamma_n)$ survives (see Figs. \ref{CV-link}b and \ref{CV-link}c). We would also like to stress that
Dirac strings are naturally avoided when the collimated non-oriented gauge configurations $A(C)$ are written as non Abelian objects which are locally Abelian. Some comments about this point are given in section \ref{nonabeliandiscussion}.

\section{Center vortex peaked wave functional} \label{wave funct}

In this section, as a preliminary step 
to account for the center vortex dominance observed in the infrared regime of lattice simulations, we shall consider a wave functional concentrated at the Cartan vector potentials of center-vortex configurations  $a(\{\gamma\})$,
 \begin{align}
    & \Psi(A)= \sum_{\{\gamma\}}\psi_{\{\gamma\}}\, \delta(A-a(\{\gamma\})) \;,  \label{sp2} \\
   & a(\{\gamma\}) = \sum_n a_{\mathscr{C}_n}(\gamma_n) \;,
\label{a9}
\end{align}
where $a_\mathscr{C}(\gamma)$ is given by Eq. \eqref{thinvortex}. The amplitudes $\psi_{\{\gamma\}}$ give the weight of a particular vortex network $\{\gamma\}$ in the Yang-Mills vacuum wave functional. 
 The sum is over the 
different  networks $\{\gamma\}$ of  lines $\gamma_n$, which include loops as well as open lines   forming the closed arrays $\mathscr{V}$ discussed in the previous section.   Furthermore, $\mathscr{C}_n\in \{\mathscr{C}_{[1]},\mathscr{C}_{[2]},...,\mathscr{C}_{[N]}\} $ is the co-weight associated with the vortex line $\gamma_n$.
For a general $\mathscr{V}$ with $I$ lines,  $N$-matching points and $M$ magnetic monopoles, the sum rule $2I=NV+2M$ must be applied\footnote{At this level, which only involves vector potentials,  it is not yet possible to associate the $M\neq 0$ sector with collimated (non-oriented) center-vortex fluxes, which will be done in section \ref{ele-mon}.}. Switching to the electric field representation
\begin{align}
    \tilde{\Psi}(E)=\int [{\mathcal D}A]\, e^{i \int d^3x\, (E, A)} \Psi(A) \;,
    \label{we}
\end{align}
the quantum state \eqref{sp2} becomes
\begin{align}
    \tilde{\Psi}(E)=
    \sum_{\{\gamma\}} \psi_{\{\gamma\}} \, e^{i\sum_n\int d^3x \,( E, a_{\mathscr{C}_n}(\gamma_n)))} \; .
    \label{we1}
\end{align}
Using the explicit form of the vortex gauge potential $a_\mathscr{C}(\gamma)$ (cf. Eq. \eqref{thinvortex}), we obtain
\begin{align}
    \int d^3x \,( E(x), a_\mathscr{C}(x,\gamma)) = \int_{\gamma} dx\cdot \Lambda^{\rm T}_\mathscr{C}(x)
    \;,
    \qquad \Lambda^{\rm T}_\mathscr{C}(x)=\int d^3\bar{x}\,D(x-\bar{x}) \, \nabla_{\bar{x}} \times (\mathscr{C},E) \label{deflambda} 
\end{align}
and the wave functional \eqref{we1} becomes 
\begin{align}
        \tilde{\Psi}(E)=\sum_{\{\gamma\}}\Psi_{\{\gamma\}}  \prod_n e^{i \int_{\gamma_n}  dx \cdot  \Lambda^{\rm T}_{\mathscr{C}_n}}(x)\;.
    \label{WF1a}
\end{align}  
For an ensemble $\{\gamma\}$ of uncorrelated lines $\gamma_n$ the weight function  $\psi_{\{\gamma\}}$ is obviously given by 
\begin{equation}
    \psi_{\{\gamma\}}=\prod_n \psi_{\gamma_n} \;,
    \label{w3}
\end{equation}
where $\psi_{\gamma_n}$ is the statistical weight of the single vortex line $\gamma_n$. Such an ensemble is realized when all  $\gamma_n$ are closed by themselves, which corresponds to a  magnetic field 
\begin{align}
   \nabla \times  a(\{\gamma\}) = 
 \,\sum_{n} \mathscr{C}_nj_{\gamma_n }.
\label{Bloop}
\end{align}
In this case, $\tilde{\Psi}(E)$ in Eq. \eqref{WF1a} is an expansion  in terms of Wilson loops computed with the dual transverse gauge fields $ \Lambda^{\rm T}_\mathscr{C}$. As is well known, the set of all Wilson loops form an (overcomplete) basis for the gauge invariant wave functionals.
For didactic reasons, let us first confine ourselves to such ensembles of oriented closed center vortex loops $\{\gamma\}$ (i.e. no magnetic monopoles, no $N$-matchings) and in addition assume that all involved loops $\gamma_n$ are associated with the same co-weight $\mathscr{C}$. The corresponding wave functional, which we denote by $\tilde{\Psi}_\mathscr{C}(E)$, follows then from eqs. \eqref{WF1a} and \eqref{w3} to be given by
\begin{align}
        \tilde{\Psi}_\mathscr{C}(E)=\sum_{\{\gamma\}}\prod_{n} \Big[\psi_{\gamma_n}\, e^{i \int_{\gamma_n}  dx \cdot  \Lambda^{\rm T}_\mathscr{C}}\Big]
        =:\sum_{\{\gamma\}}\prod_{n}
        \tilde{\Psi}_{\gamma_n}(E) \;.
    \label{WF1}
\end{align}

The studies of refs. \cite{randomsu2}, \cite{randomsu3} show that the probability amplitude $\psi_{\gamma}$ for the occurrence of a given center vortex can be modelled by the tension and stiffness.  Parameterizing a line $\gamma$ by a fictitious time $s$, which we choose as the arc-length of the trajectory $x(s)$ traced out by $\gamma$ in $R^3$, the weight for an individual center vortex line $\gamma$ can be chosen as  
\begin{equation}
 \psi_{\gamma} = \exp \left[-\int_\gamma ds \left(\frac{1}{2\kappa}\dot{u}\cdot\dot{u}+\mu  \right)\right]  \;,
\label{w4}
\end{equation} 
where a dot means the derivative with respect to $s$ and $u=\dot{x}/\sqrt{\dot{x}^2}$ is the unit tangent vector of $x(s)$. The parameters  $\mu$ and $1/\kappa$ control the effect of tension and stiffness. 

We are particularly interested in the confining phase where center vortices percolate. As we will see below, percolating center vortices require a negative $\mu$. However, for $\mu < 0$ the weight in Eq. \eqref{w4} favours infinitely long vortex lines, which would result in an unstable phase. As is known from lattice studies \cite{ELRT}, center vortices show a repulsive interaction with a proper scaling behaviour towards the continuum limit. In order to account for this interaction, we shall modify the amplitude $\psi_{\{\gamma\}}$ in Eq. \eqref{w3} according to
\begin{equation}
\psi_{\{\gamma\}} =\prod_n\psi_{\gamma_n} \, \exp \left(-\frac{\lambda_0}{2} 
\int d^3x\, \rho^2(x)   \right)  \;,
\label{rho}
    \end{equation}
where
\begin{equation}
 \rho(x) = \sum_n \int_{\gamma_n} ds_n \, \delta(x-x(s_n))  \; 
 \label{rho1}
\end{equation}
is the vortex-line density. This implements the so called excluded volume effects.
Equivalently, using 
  \begin{equation}
 e^{ -\frac{\lambda_0}{2} 
\int d^3x\, \rho^2  }  = \int [D \sigma]\, 
e^{-\frac{1}{2\lambda_0} 
\int d^3x\, \sigma^2 } e^{ i \int d^3x\, \sigma(x) \rho(x) } \;,\label{gaussianweight}
\end{equation}
we can perform the shift  $\mu \to \mu -i\sigma(x)$ in Eq. \eqref{w4}, replace $\psi_{\gamma}$ by
\begin{equation}
 \psi_{\gamma} = \exp \left[-\int_\gamma ds \left(\frac{1}{2\kappa}\dot{u}\cdot\dot{u}+\mu -i\sigma(x) \right)\right]  \;,
\label{psis}
    \end{equation} 
and at the end of the calculation integrate over the auxiliary field $\sigma$ with the Gaussian weight defined in Eq. \eqref{gaussianweight}. In the following, we will not explicitly write the functional integral over $\sigma$ nor the Gaussian weight. That is, the integration measure $\int D\sigma \exp\big [-\int \sigma^2/(2\lambda_0)\big]$ will be understood until it is explicitly carried out.  
In the next step, to handle the wave functional defined by eqs. \eqref{WF1} and \eqref{psis}, we will exploit that the sum of closed loops in D=3 can be represented by a scalar effective field theory.


\subsection{Representation of the center vortex ensemble by a scalar field theory}
\label{representationloops}
Consider first the contribution $\tilde{\Psi}_{\gamma}(E)$ \eqref{WF1} of an individual vortex line $\gamma$ to the wave functional $\tilde{\Psi}_\mathscr{C}(E)$. In the total set of vortex clusters $\{\gamma\}$, a vortex line $\gamma$ (associated with a fixed co-weight $\mathscr{C}$) occurs with arbitrary length $L$ and shape. Consider first the set of vortex lines with a fixed length $L$, fixed endpoints, $x_{\rm i}$ and $x_{\rm f}$ and fixed initial and final tangent vectors, $u_{\rm i}$ and $u_{\rm f}$,  but arbitrary shape. 
Exploiting methods from polymer physics \cite{fred} we treat these vortex lines as "wormlike chains" and represent the sum over them by a functional integral. Collecting position $x$ and unit tangent vector $u=\dot{x}/\sqrt{x^2}$ in a single letter $v=(x,u)$, the contribution of
this set lines, $\gamma(v_{\rm f},v_{\rm i},L)$, is given by
\begin{equation}
     \tilde{\Psi}[\gamma(v_{\rm f},v_{\rm i},L)](E)=\int[Dv(s)]^L_{v_{\rm f},v_{\rm i}} \, \psi_\gamma \,
    \tilde{\Psi}_{\gamma}(E) \;,
   \label{Q1}
\end{equation}
where $[Dv(s)]^L_{v_{\rm f},v_{\rm i}}$ integrates over open lines $\gamma$ with length $L$ and with initial and final coordinates  $v_{\rm i}$ and $v_{\rm f}$, respectively.
Inserting here the explicit form of the amplitude $\tilde{\Psi}_{\gamma}(E)$ \eqref{WF1}, \eqref{psis} we find 
\begin{equation}
    \tilde{\Psi}[\gamma(v_{\rm f},v_{\rm i},L)](E)=\int[Dv(s)]^L_{v_{\rm f},v_{\rm i}} \exp\left(-\int_0^Lds \,  \mathcal{L}(x(s))\right) \;,
    \label{w8}
\end{equation}
with
\begin{equation}
    \mathcal{L}(x(s))=\frac{1}{2\kappa}\dot{u}(s)\cdot\dot{u}(s)+\mu -i\sigma(x) -i\dot{u}(s)\cdot\Lambda^T(x(s),E) \;.
    \label{lag}
\end{equation}
Eq. \eqref{w8} is the (Euclidean) functional integral representation of the wave function of a particle with  classical Lagrangian \eqref{lag}. 
Accordingly it satisfies the Euclidean "Schr\"odinger equation" (with the length $L$ playing the role of the Euclidean time), which is the heat equation
\begin{align} 
   -\partial_L \tilde{\Psi}[\gamma(v,v_0,L)](E)=\left( -\frac{\kappa}{2}\Delta_u+\mu -i\sigma(x)+u\cdot(\nabla-i \Lambda^{\rm T}_\mathscr{C})\right) \tilde{\Psi}[\gamma(v,v_0,L)](E)  \;, \label{completediffusion}
\end{align}
where $\Delta_u$ is the Laplacian on the unit sphere $S^2$. As in the analogous case of the point particle this amplitude satisfies the initial condition
\[
\tilde{\Psi}[\gamma(v,v_0,L=0)](E)=\delta^{(3)}(x-x_0)\, \delta^{(2)}(u-u_0) \;.
\] 
After expanding the $u$-dependence of \eqref{completediffusion} using spherical harmonics, an infinite set of of coupled equations for the different angular momenta ($l$) can be obtained. In the limit of small stiffness $1/\kappa$, the dominant term $  \tilde{\Psi}_0[\gamma(x,x_0,L)](E)$ is $u$-independent ($l=0$) and satisfies the ordinary diffusion equation \cite{diff1,diff2,diff3,oxmanreinhardt,O-S-D-3d}
\begin{align} 
   -\partial_L \tilde{\Psi}_0[\gamma(x,x_0,L)](E)=  O_\mathscr{C} \tilde{\Psi}_0[\gamma(x,x_0,L)](E)  \;, \label{approxdiffusion}
\end{align}
\begin{align}
    O_\mathscr{C}=-\frac{1}{3\kappa}
    D^2(\Lambda_\mathscr{C}^{\rm T})+\mu -i\sigma(x) \;,\qquad D(\Lambda_\mathscr{C}^{\rm T})=\nabla-i\Lambda_\mathscr{C}^{\rm T} \;.
    \label{Qprop}
\end{align} 
Then, the amplitude \eqref{w8} becomes the usual quantum transition amplitude of a particle with Hamiltonian $O_\mathscr{C}$
\begin{align}
    \tilde{\Psi}_0[\gamma(x,x_0,L)](E) \approx \langle x |e^{-LO_\mathscr{C} }|x_0\rangle 
    \label{Qprop1}\;.
\end{align}

Consider now the set of all closed oriented center vortex loops. From a single vortex line $\gamma(v_{\rm f},v_{\rm i},L)$ we find an oriented closed loop of length $L$ by identifying the initial and final coordinates $v_{\rm f}=v_{\rm i}$ and integrate over them. 
Furthermore, we have to sum over vortex loops of arbitrary length $L$. This leads to the integral $\int_0^{\infty}dL/L...$ where the factor $1/L$ has to be included to avoid overcounting due to the choice of reference point on the loop. Finally, we have to sum over an arbitrary number $n$ of loops, which results in the sum $\sum_{n=0}^{\infty}1/n!...$ where the factor $1/n!$ is needed since a permutation of loops does not result in a new vortex configuration. Taking all this into account, we find for the wave functional generated by the set of all oriented center vortex loops associated with the co-weight $\mathscr{C}$
\begin{equation}
    \tilde{\Psi}_\mathscr{C}(E) \approx{\rm exp}\left[\int_0^\infty \frac{dL}{L}\int dx \, \tilde{\Psi}_0[\gamma(x,x,L)](E)\right]
    \label{w10}
\end{equation}
Inserting here the expression \eqref{Qprop1} and using the proper-time representation for the logarithm  
\begin{equation}
    \int_0^\infty \frac{dL}{L}\exp{(-LO)}=-\log O+ {\rm const.}
\end{equation}
as well as 
\begin{equation}
   \int dv \langle v |\log O|v\rangle ={\rm Tr}\, (\log O)=\log \det O
\end{equation}
and, furthermore, representing the functional determinant by a complex scalar field $\phi$,
we finally obtain for the wave functional \eqref{w10} \begin{equation}
    \tilde{\Psi}_\mathscr{C}(E)=\int D(\phi^\dagger, \phi) \, \exp \left[-\int d^3x \,
    \phi^\dagger O_\mathscr{C} \phi \right].
    \label{w11}
\end{equation}

So far, we have included the set of oriented vortex loops, all being associated with the same co-weight $\mathscr{C}$. Taking now into account that there are $N$ different co-weights $\mathscr{C}_{[k]}$ and each of the associated set of center vortex loops contributes a factor \eqref{w11} to the wave functional we obtain
\begin{align}
\tilde{\Psi}_0(E):=\prod\limits_{j=1}^N \tilde{\Psi}_{\mathscr{C}_{[j]}}(E) =\int \prod\limits_{j=1}^N D(\phi^\dagger_j, \phi_j) \,{\rm exp}\left[-\int d^3x\, \sum_{j=1}^N \bar{\phi}_j\, O_{\mathscr{C}_{[j]}} \phi_j \right]\;. \label{effmixvortex}
\end{align}

 Let us now also include $N$-matchings as well as magnetic monopoles. 
In this case, the amplitude $\psi_{\{\gamma\}}$ in Eq. \eqref{w3} has, of course, to be modified. We denote the probability(amplitude) that two vortex lines of the cluster $\{\gamma\}$ match at one of their endpoints to form a magnetic monopole by $\vartheta_0$ and the probability(amplitude) that $N$ vortex line form an $N$-matching point by $\xi_0$. If a vortex cluster $\{\gamma\}$ contains $V$ $N$-matching points and $M$ magnetic monopoles the weight function is then given by 
\begin{equation}
    \psi_{\{\gamma\}}=\xi_0^V\vartheta_0^M\prod_n \psi_{\gamma_n} \,.
    \label{WF3}
\end{equation}
Note that, although oriented and non-oriented center-vortices give the same result for the Wilson loop, their treatment in the
ensemble must be different. The sum over vortex clusters in Eq. \eqref{we1} includes in fact integrals over the monopole positions (see App. A). Therefore, 
the physical parameter $\vartheta_0$ is  essential to match the dimensions of contributions with different numbers of monopoles. The repulsive interaction between center vortices introduced in Eq. \eqref{rho} was observed on the lattice for $SU(2)$, where only one species of center vortices exists \cite{ELRT}. Therefore we shall assume here that this interaction occurs only between vortices of the same species. Then the shift introduced before Eq. \eqref{psis} becomes co-weight dependent, $\mu\to\mu-i\sigma_j$, and we have $N$ scalar fields $\sigma_j$, one for each co-weight $\mathscr{C}_{[j]}$, which have to be integrated with the Gaussian weight  $\frac{1}{2\lambda_0} 
\int d^3x\, \sum_j \sigma_j^2$. Extending the analysis for the oriented vortex loops to this general case (see Appendix A), one arrives at the following wave functional
\begin{align}
    &\tilde{\Psi}(E)=\int \prod\limits_{k=1}^N D(\bar{\phi}_k,\phi_k) \exp {\left[ -W[\phi;\Lambda^{\rm T}]\right]}\makebox[.4in]{,}\label{waverepr}\\ &W[\phi;\Lambda^{\rm T}] = \int d^3x\left(-\frac{1}{3\kappa} \sum_{k=1}^N\bar{\phi}_k D^2(\Lambda^T_{\mathscr{C}_{[k]}})\phi_k+V(\phi)\right)\;,\nonumber\\&
     V(\phi)  =\frac{\lambda_0}{2}\sum_k \left(\bar{\phi}_k \phi_k +\frac{\mu}{\lambda_0}\right)^2  -\xi_0 \,(\prod_{k=1}^N \phi_{k} +{\rm c.c.})-\vartheta_0\sum\limits_{k\neq l}\bar{\phi}_k\phi_{l} 
    \label{WF2}\;.
\end{align}

The upshot of the inclusion of the magnetic monopoles and $N$-matchings is the appearance of interactions between the scalar fields $\phi_k$ associated to different co-weights $\mathscr{C}_{[k]}$. This is of course expected since magnetic monopoles and $N$-matchings occur when center vortex lines associated with different co-weights match at one of their endpoints.

For later use let us discuss the symmetry of the action $W(\phi;\Lambda)$ of the effective field theory \eqref{WF2}.
If only oriented vortex loops were included (i.e. neglecting N-matching and magnetic monopoles), which corresponds to $\xi_0=0, \vartheta_0=0$, the theory is obviously invariant with respect to a separate change of the phase of the individual fields $\phi_k, $
\begin{equation}
    \phi_k\to e^{i\varphi_k}\phi_k,\qquad k=1,...,N
    \label{s1}
\end{equation}
so it has an $U(1)^N$ symmetry. When we include $N$-matching, $\xi_0\neq 0$ but still exclude magnetic monopoles, $\vartheta_0=0$, the phases of the individual fields can no longer be independently changed but have to satisfy the constraint
\begin{equation}
   e^{\varphi_1}e^{\varphi_2}...e^{\varphi_N}=1.
   \label{s2}
\end{equation}
in order to keep the potential invariant, i.e. $N-1$ phases $\varphi_i$ can be chosen independently and the remaining phase is then determined by Eq. \eqref{s2}. This condition reduces the symmetry to $U(1)^{N-1}$.
When magnetic monopoles, which arise from the matching of the endpoints of two vortex lines $\gamma_k$ and $\gamma_l$ associated with different co-weights $\mathscr{C}_k\neq\mathscr{C}_l$, are included, $\vartheta_0 \neq 0$, but $N$-matchings are excluded, $\xi_0=0$, invariance of the potential requires the constraints 
\begin{equation}
   e^{i\varphi_k}=e^{i\varphi_l}
   \label{s3}
\end{equation}
for each pair of co-weights. The potential is then only invariant with respect to a simultaneous change of the phases of all fields by the same amount
\begin{equation}
    \phi_k\to \exp{i\varphi}\phi_k,\qquad k=1,...,N
    \label{s4}
\end{equation}
and  as a consequence the symmetry of the classical potential is reduced to $U(1)$. 
Finally, including magnetic monopoles and $N$-matching points, $\xi_0\neq0, \vartheta_0\neq0$, both constraints \eqref{s2} and \eqref{s3} have to be fulfilled. This   
restricts the possible phase transformations in Eq. \eqref{s4} to
those satisfying the condition
\begin{equation}
    \exp{iN\varphi}=1 \qquad \to \qquad \varphi=n2\pi/N, \qquad n=0,1,2,...,N-1
    \label{s5}
\end{equation}
and the theory is only invariant with respect to a multiplication of all fields $\phi_k$ by the center elements:
\begin{equation}
    \phi_k \, \to \, Z_n \phi_k,\qquad Z_n=\exp{(in2\pi/N)},
\end{equation}
leaving the symmetry group $Z(N)$. Accordingly, the vacuum field configurations of scalar fields $\phi_k$ will be characterized by a center element, see section \ref{sectionwilson}.

\subsection{Collimating the non-oriented center vortex component}
\label{ele-mon}

We are eventually interested in calculating the Wilson loop average. Since the wave functional constructed above has support only on Cartan gauge potentials $a(\{\gamma\})$, we can use the ordinary Stokes theorem to express the Wilson loop as
\begin{equation}
    W\left[a\right](C)  =\frac{1}{N}{\rm Tr}\, \left({\rm exp}\left[i\int_{S(C)} dS \cdot B \right]\right) \;,
    \label{WL}
\end{equation}
$B= \nabla \times a$, where $S(C)$ is an arbitrary area bounded by the loop $C$. 
As discussed in section \ref{anatomy}, the curl of the Cartan gauge field associated to chains contains collimated (vortex) and non-collimated (monopole-like) fluxes (see e.g. Eq. \eqref{m2}). 
Accordingly, we obtain two multiplicative contributions to the Wilson loop
\begin{equation}
     W\left[a\right](C)  = W_{\rm coll} \, W_{\rm non-coll} \;,
    \label{b0}
\end{equation}
where $W_{\rm coll}$ yields a center element. That is, there is an extra monopole-like contribution to the Wilson loop, in addition to the center element produced by center-vortex configurations (cf. Eq. \eqref{wl2}). In order to comply with the collimated flux property of Abelian projected chain configurations observed in the lattice,
we introduce a dual Cartan scalar potential $\zeta$ and consider the replacement
\begin{equation}
 B=\nabla \times a(\{\gamma\}) -\nabla \zeta
   \label{b1}
\end{equation}
in Eq. \eqref{WL}, such that the total flux $B$
only contains the collimated part. This way, the Wilson loop becomes a pure center element as in the case of the center projected lattice. This can also be thought of as getting rid of unphysical Dirac strings (see Eq. \eqref{ds}), leaving only the physical effect originated from their endpoints. 
From eqs. \eqref{a3} and \eqref{a9}, this requires
\begin{equation}
    \zeta(x)=(-\Delta)^{-1}\nabla \cdot b(x,\{\gamma\}) \makebox[.5in]{,} b(x,\{\gamma\}) = \sum_n \mathscr{C}_n j(x,\gamma_n) \;,
     \label{b2}
\end{equation}
where the sum runs over all vortex lines $\gamma_n$ forming the vortex cluster $\{\gamma\}$. Then, for a general   $\{\gamma\}$, the total flux becomes
\begin{equation}
    B = b(x,\{\gamma\}) \;.
     \label{b3}
\end{equation}
Accordingly, the modified Wilson loop now yields 
$ Z^{L(\{\gamma\},C)} $,
where $L(\{\gamma\},C)$ is the linking number between the Wilson loop and the vortex cluster $\{\gamma\}$. Of course, 
this would also be obtained if the Dirac strings in Eq. \eqref{ds} were added to the gauge field configuration $a(\mathcal{V})$ in Eq. \eqref{a7}.
 
With the introduction of the dual scalar potential $\zeta$ constrained by Eq. \eqref{b2}, our vortex wave functional becomes
\begin{equation}
   \Psi (A,\zeta)=\sum_{\{\gamma\}}\psi_{\{\gamma\}}\, \delta\big(A-a(\{\gamma\})\big)\, \delta\big(\zeta-(-\Delta)^{-1}\nabla b(\{\gamma\})\big) \;,
   \label{b4}
\end{equation}
where $a(\{\gamma\})$ and $b(\{\gamma\})$ are given by eqs. \eqref{a9} and \eqref{b3}, respectively.
Analogously to the electric field representation \eqref{we}, we define a dual representation for the wave functional \eqref{b4} by 
\begin{align}
    \tilde{\Psi}(E, \eta)=\int [DA]\int [D\zeta]\, e^{i\int d^3x (E,A)}e^{i\int d^3x (\zeta,\eta)}\Psi(A,\zeta)\;.
    \label{WFD}
\end{align}
Inserting here the explicit form of our vortex wave functional \eqref{WF1}, we obtain
\begin{equation}
\tilde{\Psi}(E,\eta)= \sum_{\{\gamma\}} \psi_{\{\gamma\}}\exp {\left(i\sum_{n}\int_{\gamma_n} dx\cdot \Lambda_{\mathscr{C}_{n}}(E,\eta) \right)}\;,
\label{WF4}
\end{equation}
which is the same expression as the original electric field representation \eqref{WF2} except that the transverse field $\Lambda^T(E)$ \eqref{deflambda} is replaced by
\begin{equation}
   \Lambda_{\mathscr{C}}(E,\eta)=\Lambda_{\mathscr{C}}^{\rm T}(E)+\Lambda_{\mathscr{C}}^{\rm L}(\eta)\;,
\label{ll2} 
\end{equation}
whose longitudinal part is 
\begin{equation}
\Lambda^{\rm L}_{\mathscr{C}}(x,\eta)=\int d^3\bar{x}\, D(x-\bar{x})\nabla_{\bar{x}}(\mathscr{C},\eta)\;. 
\label{ll3}
\end{equation}
Now, repeating the steps that led to the effective field theory description \eqref{waverepr}, \eqref{WF2} of the center-vortex ensemble, we find for  \eqref{WF4} 
\begin{align}
    &\tilde{\Psi}(E,\eta)=\prod\limits_{j=1}^N\int D(\bar{\phi}_k,\phi_k) \exp \left[ -W[\phi,\Lambda]\right]
    \label{ww1}\;.
\end{align}
Now, with Eq. \eqref{b4}, the scalar product in the Hilbert space of the Yang-Mills wave functional also includes a functional integration over the scalar field $\zeta$ and the expectation value of the Wilson loop \eqref{WL}, \eqref{b1} becomes
 \begin{align}
    &\langle W_{\rm D}(C)\rangle =\frac{1}{\mathscr{D}}\int [DA][D\zeta]\, {\rm Tr}\,{\rm D}\left({\rm exp}\left[i\int_{S(C)} dS \cdot (\nabla\times A-\nabla\zeta) \right]\right)|\Psi(A,\zeta)|^2\;.
    \label{wl3}
\end{align}
Here, we considered a general $\mathscr{D}-$dimensional quark representation $D(\cdot)$ of $SU(N)$. Defining the (vector-valued) characteristic function of the area $S(C)$ bounded by the Wilson loop $C$ 
\begin{equation}
    \Sigma(x,S(C))=\frac{1}{2}\int_{S(C)} d\sigma_1d\sigma_2\,\frac{\partial x}{\partial \sigma_1}\times\frac{\partial x}{\partial \sigma_2} \, \delta\big(x-\bar{x}(\sigma)\big) \;,
    \label{wwc}
\end{equation}
with $\bar{x}(\sigma)$ being a parametrization of $S(C)$
\footnote{Note that the characteristic function satisfies  $\oint_{C'} dx\cdot \Sigma(x,S(C)) = I(S(C),C')=L(C,C')$
 where $I(S(\mathcal{C}),C')$ is the intersection number between the area $S(C)$ and the loop $C'$.},
we find
\begin{align}
 \langle W_{\rm D}(C) \rangle =\frac{1}{\mathscr{D}}\sum_\Omega\int [DA][D\zeta]\,\left({\rm exp}\left[-i\int d^3x \, (A^q\cdot\nabla\times \Sigma-\zeta^q \nabla\cdot \Sigma)\, \Omega^q\right]\right)|\Psi(A,\zeta)|^2\;.
 \label{wl4}
  \end{align}
In the above expression, the trace was calculated in the basis in which the Cartan generators ${\rm D}(T_q)$ are diagonal. The tuple of eigenvalues $\vec{\Omega} = (\Omega^1, \dots, \Omega^N)$ for a given common eigenvector, i.e. ${\rm D}(T_q) |\Omega\rangle = \Omega^q |\Omega\rangle $, are the  weights of the representation. In the dual representation \eqref{WFD}, this expectation value becomes the  convolution
\begin{align}
       \langle W_{\rm D}(C) \rangle =\frac{1}{\mathscr{D}}\sum_{\Omega}\int [DE][D\eta]\,\tilde{\Psi}^*(E,\eta)\,\tilde{\Psi}(E+\Omega\, \nabla\times \Sigma(S),\eta-\Omega\, \nabla\cdot \Sigma(S))\;,
    \label{wl5}
\end{align}
where $\Omega=\Omega^qT_q$. 
\subsection{Non-Abelian representation of collimated vortex configurations}\label{nonabeliandiscussion}
\begin{figure}[h]        
\centering 
\includegraphics[scale=0.4]{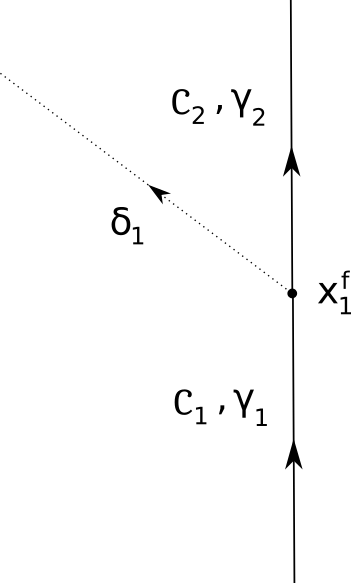}
\caption{\small The collimated flux associated with the Abelianized configuration  $a_{\mathscr{C}_{1}}(\gamma_1)+a_{\mathscr{C}_2}(\gamma_2) + a_{\mathscr{E}_1}(\delta_1)  $. Besides the physical contributions $\mathscr{C}_{1}$ ($\mathscr{C}_{2}$) carried by $\gamma_1$ ($\gamma_2$), we display the unobservable Dirac string $\delta_1$, which carries flux $\mathscr{C}_{1}-\mathscr{C}_{2}$. The arrows give the orientation of the lines.}    
\label{CD} \end{figure}

Before moving to section \ref{sectionwilson}, where we estimate the expectation value of the Wilson loop, we would like to discuss how   collimated configurations are accomodated in the Yang-Mills  context. In fact, not only center-vortex loops and arrays with $N$-matching, but also collimated fluxes formed by non-oriented components can be represented in terms of  non-Abelian  gauge fields.  
All of them can be written in the form \cite{diff1}, \cite{conf-qg}
 \begin{gather} 
 {\rm Ad}(A_{\rm coll}) = 
i R\nabla R^{-1}    \makebox[.5in]{,}    R = {\rm Ad}(S)  \makebox[.5in]{,} S \in SU(N)\;, \label{defvort}
\end{gather} 
where $S$ changes by a center element when going around the center vortices.\footnote{ 
The use of  ${\rm Ad}(S)$ is equivalent to subtract a contribution localized on a surface (ideal 
center-vortex),  after computing $ i\,  S \partial_i S^{-1} $, as done in Ref. \cite{reinhardtvortices}.} They are thin center-vortex configurations in the sense that the Wilson loop computed with 
$A_{\rm coll}$ always gives a center element (cf. Eq. \eqref{wl2}), while the field-strength is always localized at the center-vortex defects, where $R(S)$ is multivalued,
\begin{gather}
   {\rm Ad} (F_{ij}) = i R [\partial_i , \partial_j] R^{-1} \;.
\end{gather}
For example, for Abelian center-vortex loops and lines with $N$-matching,  $S$ is in the Cartan subgroup, which gives $A_{\rm coll}= \sum_n a_{\mathscr{C}_n}(\gamma_n)$.   On the other hand, for non-oriented configurations,  $S=V W$, where $V$ is Cartan and changes by a center element when going around the vortices, while $W$ is single-valued and changes the orientation of the flux. This can be better visualized 
by writing $A_{\rm coll}$ in terms of a local Lie basis 
\[
n_A =S T_A S^{-1} \makebox[.5in]{,} A=1, \dots, N^2-1 \;,
\]
where $n_q$, $q=1, \dots, N-1$ are local Cartan directions ($[n_q, n_p]=0$). The off-diagonal generators can be labelled by $N(N-1)$ tuples $\vec{\alpha}$ formed by $N-1$ components $\alpha^q$. 
For each $\vec{\alpha}$, there is a pair $\{ T_\alpha, T_{\bar{\alpha}}\}$ that together with $ \alpha = \alpha^q T_q$ generate an $\mathfrak{su}(2)$ subalgebra of $\mathfrak{su}(N)$ \footnote{For $N=2$,  the pairs $\{ T_\alpha, T_{\bar{\alpha}}\}$ are in correspondence with the Pauli matrices in $\{ \sigma_1, \sigma_2\} $, while for $N=3$, they correspond to the Gell-Mann matrices in $\{ \lambda_1$, $\lambda_2\}$, $\{ \lambda_4$, $\lambda_5\}$, and $\{\lambda_6$, $\lambda_7\}$.},
\begin{gather}
[\alpha, T_\alpha] = \frac{i}{N} \, T_{\bar{\alpha}}    \makebox[.5in]{,}    [T_\alpha , T_{\bar{\alpha}}] = i\, \alpha \;.
\end{gather}
When the local Cartan directions contain point-like defects, the local off-diagonal directions $n_\alpha$, $n_{\bar{\alpha}}$  contain 
defects localized on lines. Take for example \cite{conf-qg},
\begin{equation}
S= \exp \left(  i \frac{\varphi}{2\pi}\, \mathscr{C}_2\right)\, W(\theta)
\makebox[.5in]{,}
W(\theta)=\exp \left( i\theta \,\sqrt{N} T_{\alpha} \right) \;,
\label{ele-comp}
\end{equation}
where $\varphi$ and $\theta$ are the polar angles, $ \mathscr{C}_2 = 2\pi 2N \omega_2 $, and $T_\alpha$ is labelled 
by the root $\vec{\alpha} = \vec{\omega}_1 - \vec{\omega}_2$. In this case
\begin{gather}
A_{\rm coll}= S \left[a_{\mathscr{C}_{1}}(\gamma_1)+a_{\mathscr{C}_2}(\gamma_2) \right]S^{-1}
+i\, [L_{\alpha}, \nabla L_{\alpha}]   \;,
\label{ex-coll}
\end{gather} 
\begin{gather}
L_{\alpha}= S N\alpha S^{-1} =  \cos \theta\, N\alpha + \sin \theta \cos 
\varphi \, \sqrt{N} T_{\bar{\alpha}} + \sin \theta \sin \varphi 
\, \sqrt{N} T_{\alpha}\;, 
\end{gather} 
where $\gamma_1$, $\gamma_2$ are lines running along the $z$-axis (see Fig. \ref{CD}). 
 Because of $W(\theta)$, $L_\alpha$ is a topologically nontrivial map from $S^2$, parametrized by $\theta, \phi$,  into vectors in an $\mathfrak{su}(2)$ subalgebra of $\mathfrak{su}(N)$. Consequently, it is not possible to perform a regular gauge transformation so as to align the gauge field along the global Cartan directions $T_q$. Nevertheless,
we can embedded this configuration in the lattice and determine the maximal Abelian gauge form for the associated link-variables.
 To do so, we can consider the mapping $ S_{\rm D}$ 
  \begin{gather}
 S_{\rm D}\, \alpha\, S_{\rm D}^{-1} =  S\, \alpha \, S^{-1} \makebox[.5in]{,}    S_{\rm D} = S \exp \left( - i \frac{\varphi}{2\pi}\, \mathscr{C}_2\right) \, \exp \left(  i \frac{\chi}{2\pi}\, (\mathscr{C}_2 - \mathscr{C}_1 )\right) \;,
 \end{gather}
 where $\chi$ changes by $2\pi$ when going around the path $\gamma_1 \cup \delta_1$ (see Fig. \ref{CD}) in the positive sense. It leads to the same local Cartan directions $L_\alpha$ than $S$ in Eq. \eqref{ele-comp} and satisfies 
 \begin{align}
i S_{\rm D} \nabla S_{\rm D}^{-1} = -S_{\rm D} \, a_{\mathscr{E}_1}(\delta_1) \, S_{\rm D}^{-1} +i\, [L_{\alpha}, \nabla L_{\alpha}] \makebox[.5in]{,} \mathscr{E}_1 = \mathscr{C}_1 - \mathscr{C}_2  \;. 
\end{align}
Note that the left-hand side of this equation can be written without relying on the adjoint representation (cf. \eqref{defvort}) because $S_{\rm D}$ is single-valued when going around any loop. This, together with Eq. \eqref{ex-coll}, yields
  \begin{align}
A_{\rm coll} = S_{\rm D} \left( a_{\mathscr{C}_{1}}(\gamma_1)+a_{\mathscr{C}_2}(\gamma_2) + a_{\mathscr{E}_1}(\delta_1)  \right)\, S_{\rm D}^{-1} 
+ i S_{\rm D} \nabla S_{\rm D}^{-1}\;.
\end{align} 
Although $S_{\rm D}$ has (Dirac string) defects, the calculation of any Wilson loop for $A_{\rm coll}$ and $a_{\mathscr{C}_{1}}(\gamma_1)+a_{\mathscr{C}_2}(\gamma_2) + a_{\mathscr{E}_1}(\delta_1)  $ gives the same result. Then, when embedded in the lattice, the corresponding link-variables become equivalent.
This is because $S_{\rm D}$ leads to a well-defined field on the lattice sites, as long as the Dirac strings do not pass through these points.  In other words, the lattice maximal Abelian gauge applied to the link-variables for  $A_{\rm coll}$ would agree with the  collimated Abelian fluxes (with the additional trivial plaquettes) observed in the lattice and modelled throughout this work (see section \ref{Abep}). In Fig. \ref{CD}, we show the flux for the Abelianized field $a_{\mathscr{C}_{1}}(\gamma_1)+a_{\mathscr{C}_2}(\gamma_2) + a_{\mathscr{E}_1}(\delta_1)$ in the continuum. This illustrates the situation in Figs. \ref{CV-link}b and \ref{CV-link}c around $x^{\rm f}_1$. 

It is interesting to note that the thin collimated configurations can be thickened and  
accommodated in a set of restricted gauge fields. The latter were introduced for $SU(2)$ in Ref. \cite{refcho}, and generalized to $SU(N)$ in Refs. \cite{gen1,conf-qg}. 
These restricted fields are ``locally'' Abelian configurations in the sense that, in regions with no ideal center vortices, 
they can be  written as the gauge transformation of a Cartan gauge field. 
In addition, the corresponding non-Abelian field strength points along the local Cartan directions $n^q$ and receives the contribution of a topological 
monopole flux. In the case of $A_{\rm coll}$, this term is responsible for the 
flux collimation (see the discussion in Refs. \cite{monopoles1}, \cite{conf-qg}). 
 We also note that
 all possible smooth non-Abelian vector gauge fields 
can be  separated into sectors labelled by $SU(N)$ mappings with defects \cite{singer1, singer2}. This allowed for the implementation of a sector-dependent gauge fixing procedure that could circumvent Singer's no-go theorem. 

\section{The average of the Wilson Loop operator}\label{sectionwilson}
The Wilson loop is an important order parameter for confinement, which has been intensively studied both on the lattice and in the continuum. Its expectation value provides information about the potential  between static sources. At asymptotic distances, the string tension $\sigma_{\rm D}$ is known to depend only on the $N$-ality $k$ of the representation ${\rm D}(\cdot)$, defined by the relation
\begin{align}
    {\rm D}(e^{i\frac{2\pi}{N}}I)=e^{i\frac{2\pi k}{N}} I\;.
\end{align}
Moreover, due to gluon screening, among the strings of representations  with a given $N$-ality $k$, only the one associated to the smallest string tension will be stable. As for the precise dependence of the string tension with $k$, current lattice data cannot distinguish between a Casimir and a Sine law,
\begin{align}
\sigma_k^{\rm Casimir}=\sigma\frac{k(N-k)}{N-1}    \makebox[.5in]{,}   \sigma_k^{\rm Sine}=\sigma \frac{\sin(k\pi/N)}{\sin(\pi/N)} \;,
\label{sin}
\end{align}
where $\sigma$ is the fundamental string tension. As discussed in Ref. \cite{teper}, for both the Sine and Casimir scenarios, the most stable strings are those of the $k$-Antisymmetric representation, which will be considered from now on.

To elaborate the expression for the Wilson loop \eqref{wl5} derived above and also to exploit results obtained in previous work we rewrite  the action  of the effective field theory in a more compact form. The fields $ \phi_k$ appearing in the vortex wave-functional \eqref{ww1} can be assembled into an $ N\times N$ matrix
\begin{align}
    \Phi=\frac{1}{\sqrt{3\kappa}}\begin{pmatrix}
\phi_1 & 0 & 0 &  \dots \\
0 & \phi_2 & 0 & \dots  \\
0 &0 & \phi_3 & \dots \\
\vdots & \vdots &\vdots &\ddots
\end{pmatrix}. 
\label{ww2}
\end{align}
In addition, the Fierz identity 
\footnote{Note that the summation over $A$ runs here over all generators, not just over the generators of the Cartan group.}
\begin{align}
    T^A_{ij}T^A_{kl}=\frac{1}{2N}\left(\delta_{il}\delta_{jk}-\frac{1}{N}\delta_{ij}\delta_{kl}\right)\;,
    \label{ww3}
\end{align}
which holds for the normalization ${\rm Tr}\, (T_AT_B)=\frac{\delta_{AB}}{2N}$, implies
\begin{align}
    \sum\limits_{i,j}\bar{\phi}_iT_A^{ij}T_A^{ji}\phi_{j}=\frac{1}{2N}\sum\limits_{i,j}\bar{\phi}_i\phi_{j}\left(1-\frac{1}{N}\delta_{ij}\right)\;.
    \label{ww4}
\end{align}
Using these relations, the action \eqref{ww1}  can be rewritten as
\begin{align}
   &
   W( \Phi;\Lambda)=\int d^3x\left({\rm Tr}((D(\Lambda)\Phi)^\dagger D(\Lambda)\Phi)+V( \Phi)\right)\nonumber\makebox[.5in]{,}D=\nabla-i\Lambda\;,\\&
    V( \Phi)=\frac{\lambda}{2}{\rm Tr}(\Phi^\dagger\Phi-a^2 I_N)^2-\xi({\rm det}\Phi+{\rm det}\Phi^\dagger)-\vartheta {\rm Tr}(\Phi^\dagger T_A\Phi T_A)\;,
    \label{ww5}
\end{align}
where we introduced the Lie-algebra valued field 
\begin{align}
    &\Lambda=\Lambda^T+\Lambda^L\;, \nonumber\\&\Lambda^T=2\pi2N\int d^3\bar{x} \, D(x-\bar{x})\,\nabla_{\bar{x}}\times E\makebox[.5in]{,}\Lambda^L=2\pi2N\int d^3\bar{x}\,D(x-\bar{x})\,\nabla_{\bar{x}}\eta\;.
\end{align}
It is clear that the columns of $\Phi$ are proportional to the weight vectors of the defining representation, whose $i$-th entry equals one, while the rest are zero. Therefore, the $i$-th column of $D(\Lambda)\Phi$ is
\begin{equation}
    (D(\Lambda)\Phi)|_i= (3\kappa)^{-1/2} D(\Lambda_{\mathscr{C}_{[i]}})\phi_i\;,
\end{equation}
which makes contact with the scalar derivative in Eq. \eqref{WF2}. Moreover, the parameters in the potential \eqref{ww5} are related to those in Eq. \eqref{WF2} by
\begin{eqnarray}
&&    \lambda=9\kappa^2\lambda_0
    \makebox[.5in]{,} a^2=-\frac{\mu}{3\kappa\lambda_0}-\frac{ \vartheta_0}{3\kappa\lambda_0}\frac{N-1}{N}\;,\nonumber\\
    && \xi=(3\kappa)^{\frac{N}{2}}\xi_0 \makebox[.5in]{,} \vartheta=6\kappa N \vartheta_0\;.
    \label{para}
\end{eqnarray} 
 In this compact representation the $Z(N)$-symmetry of the potential is manifest, see also the discussion at the end of section \ref{representationloops}. 

To find the Wilson loop from Eq. \eqref{wl5}, we have to calculate the dual wave functional $\tilde{\Psi}$ (without and with displaced arguments), which is given by the scalar field theory \eqref{ww5}. Let us first consider the case with undisplaced arguments, $\tilde{\Psi}(E,\eta)$. 
For sufficiently large $\lambda_0$, i.e. sufficiently strong vortex interaction, the saddle points are approximately given by the minima of the potential, which should be chosen as space independent in order to minimize the action. In the percolating regime $(\mu<0)$, the scalar field develops a non-zero vacuum value: the minima of the potential occur at the field configurations characterized by a center element
\begin{align}
\Phi_n=v Z_n I_N ,\qquad Z_n=e^{i\frac{2\pi n}{N}},\qquad n=0,1,2,...,N-1
\;,\label{vev}
\end{align}
where $I_N$ is the $N$-dimensional unit matrix. Furthermore, the vacuum value $v$ of the field $\Phi$ is obtained by minimizing the potential \eqref{ww5} for the ansatz \eqref{vev} w.r.t. $v$, resulting in the equation 
\begin{align}
    2\lambda N(v^2-a^2)-2\xi Nv^{N-2}-\vartheta\frac{N^2-1}{N}=0
\;.\label{vev1}
\end{align}
To lowest order in the corresponding saddle-point approximation (replacing the integral by its integrand at the saddle-point) the wave functional \eqref{ww1} is then given by
\begin{equation}
    \tilde{\Psi}(E,\eta) \approx \exp \left[ -W(vI_N;\Lambda)\right]\makebox[.4in]{,} W(vI_N;\Lambda) =v^2 \int d^3x \,{\rm Tr}(\Lambda^2),
    \label{ww7}
\end{equation}
Since $\Lambda$ is linear in the dual variables $E,\eta$, Eq. \eqref{ww7} gives a Gaussian wave functional peaked at $E=0$, $\eta=0$, and for sufficiently large vacuum values $v$ the fluctuations in the dual variables $E$, $\eta$ become suppressed. In the spirit of the leading order saddle-point approximation, we can then also replace the integral over $E, \eta$ in Eq. \eqref{ww1} by its integrand at $E=0$, $\eta=0$, thus obtaining for the Wilson loop \eqref{wl5} 
\begin{align}
       \langle W_{\rm D}(C)\rangle \approx {\rm const.} \sum_{\Omega}\,\tilde{\Psi}(\Omega\, \nabla\times \Sigma(S),-\Omega\,\nabla\cdot \Sigma(S))\;.
       \label{ww8}
\end{align}
Each term of the sum in Eq. \eqref{ww8} is given by a field theory in the presence of the external vector field $2\pi2N\Omega\,\Sigma(S)$. The average of the Wilson loop may then be approximated by a sum over independent saddle-points 
\begin{align}
    \langle W_{\rm D}(C)\rangle \approx {\rm const.}  \sum_{\Omega}\,\exp\left[-W(\Phi_0^\Omega;2\pi2N\Omega\,\Sigma(S))\right]\;, \label{eqsaddle}
\end{align}
where $\Phi_0^\Omega$ is the classical solution associated to the weight $\Omega$. For simplicity, in the following we will consider weights of the defining representation. The general case of an arbitrary weight and arbitrary $k$-Antisymmetric representation with $N$-ality $k=2,...,N-1$ will be treated in Appendix B. To obtain the classical solutions, we need to understand the implications of the presence of the external vector field $2\pi2N\Omega\,\Sigma(S)$, which for the defining representation is simply $\mathscr{C}\,\Sigma(S)$.
Its cancellation induces a soliton-like saddle-point. The transition between a pair of discrete vacua is localized around the minimal surface with boundary $C$. For definiteness, we shall consider a planar circular loop $C$ of radius $R$ located in the $x-y$ plane and centered at the origin. Choosing $S(C)$ for example as the complement in the x-y plane of the disc encircled by $C$, the vector field $\mathscr{C}\,\Sigma(S)$ is directed along the $z$-axis and is non-vanishing only on the surface $S(C)$ located at $z=0$. It is then not difficult to show \cite{O-S-D-3d} that the only effect of this source is to impose the boundary condition
\begin{align}
    \Phi(x,y,z\to\infty)=ve^{i\mathscr{C}}I_N
    \makebox[.5in]{,}\Phi(x,y,z\to -\infty)= v I_N\;,\qquad x^2+y^2 \le R^2.
    \label{bc}
\end{align}
and that for large loops $C$, ignoring boundary effects (i.e. neglecting gradients in the $x-y-$ directions), the action\eqref{ww5} of the soliton reduces to
\begin{align}
    W[\Phi, \Lambda]\approx \sigma A \makebox[.5in]{,}\sigma = \int dx_3 \left({\rm Tr}\left(\partial_{x_3}\Phi^\dagger\partial_{x_3}\Phi\right) +V(\Phi,\Phi^\dagger)\right)\;,
    \label{ww11}
\end{align}
where $A$ is the area of the disc enclosed by $C$. The soliton is then found by solving the one-dimensional field equation
\begin{align}
    \partial_{z}^2\Phi = \lambda\Phi(\Phi^\dagger\Phi-a^2 I_N)-\xi (\Phi^\dagger)^{-1}{\rm det}\Phi^\dagger-\vartheta T_A\Phi T_A\;,
    \label{ww12}
\end{align}
with the boundary condition \eqref{bc}. For the defining representation, we have $e^{i\mathscr{C}}=e^{-i2\pi/N}\;, \forall \mathscr{C}$. Therefore the boundary condition is the same for all co-weights.
 As discussed in Ref. \cite{O-S-D-3d}, the field equation can be solved by the Ansatz (see Appendix B for details)
\begin{align}
    \Phi=\left(\eta I_N+\frac{\eta_0}{2\pi}\mathscr{C}\right) e^{i\mathscr{C} \theta/2\pi}e^{i\alpha}
     \label{ww13}\;,
\end{align}
where the boundary condition \eqref{bc}
imposes the following constraints  to the profile functions
\begin{align}
    &\eta(-\infty)=\eta(\infty)=v\;, & \eta_0(-\infty)=\eta_0(\infty)=0\;.
\end{align}
Due to the relation
$e^{i\mathscr{C}}=e^{-i\frac{2\pi}{N}}$, the transition between the different vacua at $z\to -\infty$ and $z\to \infty$  can be made by a change of either $\theta$ or $\alpha$. As discussed in Ref. \cite{O-S-D-3d}, for the region of parameter space that implements the appropriate hierarchy of Spontaneous Symmetry Breaking, the profiles $\eta, \eta_0, \alpha$ remain essentially constant at their vacuum values ($\eta=v, \eta_0=0, \alpha=0$). 
The boundary conditions \eqref{bc} will then be accomplished by a variation of $\theta$, i.e.
\begin{equation}
    \theta(-\infty)=0,\qquad\theta(+\infty)=2\pi.
\end{equation}
Moreover, the variation of $\theta$ will be governed by the Sine-Gordon equation 
\begin{align}
    \partial^2_{z}\theta=\frac{\vartheta}{2}\sin \theta\;.\label{sinegordon}
\end{align}
Finally, to evaluate \eqref{ww11} we used Derrick's theorem, which implies that the kinetic and potential contributions are equal. Then we obtain the following approximate expression for the string tension
\begin{equation}
    \sigma=2v^2\frac{N-1}{N}\int dz\,(\partial_{z}\theta)^2.
\end{equation}
This string tension is determined by the two quantities $v$ and $\vartheta$. The first one, the vacuum value $v$ of the module of the scalar field $\Phi$, is a measure for the density of center vortex flux lines in the  Yang-Mills vacuum. The second one, $\vartheta$, enters the equation of motion for the soliton $\theta(z)$ and gives the weight (probability amplitude) of the non-oriented center vortex configurations (which contain magnetic monopoles) in the vacuum wave functional, see Eq. \eqref{WF3}. 

Above we have considered the Wilson loop for gauge fields in the defining representation, which has an $N$-ality $k=1$. The general case is worked out in the Appendix B with the following result: the string tension for a representation with $N$-ality $k\neq 1$ is related to that with $k=1$ by
\begin{align}
    \sigma_k=\frac{k(N-k)}{N-1}\sigma, \qquad k=1,2,...,N-1.
    \label{resultk}
\end{align}
Thus, in the asymptotic regime, we find for the Wilson loop an area law
with Casimir scaling,
\begin{align}
\langle W_{\rm D}(C)\rangle \approx   \exp \left(-\sigma \, 
  \frac{k(N-k)}{N-1}   A \right) \;. \label{casimirlaw} 
\end{align}
This is one of the behaviors extracted in Ref. \cite{4dlaw} from the lattice data, which cannot distinguish between this behavior and the asymptotic sine law \eqref{sin}. 


\section{Conclusions}
\label{conclusion}

In this work, we have proposed a vacuum wave functional  peaked on an ensemble of collimated center vortices to describe the deep infrared properties of $SU(N)$ Yang-Mills theory within the Hamiltonian approach. The fluxes of the center-vortex fields entering the wave functional are fixed-time counterparts of the two-dimensional vortex surfaces found on the four-dimensional lattice. The ensemble consists of oriented and non-oriented vortices, with the possibility of matching $N$ elementary vortex lines that carry $N$ different  (defining) weights of $SU(N)$. 
As shown in Ref. \cite{reinhardtvortices}, and also found on the lattice \cite{topologicallattice}, non-oriented center vortices are absolutely necessary for a non-vanishing Pontryagin index. Furthermore, in 4d ensembles of percolating center vortices, the coexistence of oriented and non-oriented components, $N$-matching rules among center-vortex surfaces, and natural matching rules among monopole lines, is essential to generate a confining flux tube \cite{diff1} (see also \cite{{universe7080253}}). Indeed, the 
center-vortex field configurations in our vacuum wave functional incorporate all the features and correlations of center vortices observed for $SU(2)$ in the indirect maximal center gauge, naturally extended to $SU(N)$. In particular, 
the change of vortex orientation (in the Cartan subalgebra) is caused by magnetic monopoles. In the   Abelian projected scenario, to describe properly the observed collimation of  non-oriented fluxes, the Cartan gauge fields associated with the center-vortex lines were supplemented by a Cartan scalar field.
The center vortices were then endowed with stiffness and, using techniques from Polymer physics, we were able to express the electric-field representation of our wave functional as an effective theory of $N$ complex scalar fields. When both oriented and non-oriented vortices as well as $N$-vortex matchings are included, the effective potential of the scalar fields has a $Z(N)$ symmetry, which is, however, broken by its vacuum configurations, given by the $N$ different center elements of $SU(N)$. 
Using this representation of our wave functional, and relying on a saddle-point approximation  to the functional integral over the effective scalar fields, we have calculated the Wilson loop in the $k$-antisymmetric representation. The saddle-point is given by a solitonic field configuration, which interpolates between two different minima and which is localized on the minimal surface spanned by the Wilson loop. We found an area law for the Wilson loop and a string tension that shows an asymptotic Casimir scaling, which is in line with one of the possible scalings seen in lattice calculations. These properties agree with those found in the 4d ensemble of percolating center-vortex surfaces with oriented and non-oriented components (for a review, see \cite{{universe7080253}}). 
The results obtained in this work provide further evidence that the coexistence of these components, together with their natural correlations, are essential to describe all the asymptotic confining properties in Yang-Mills theory. 
In the future, we plan to use the wave functional constructed in the present paper to calculate the t'Hooft loop and the topological susceptibility.

\section*{Acknowledgments}
The Deutscher Akademischer Austauschdienst (DAAD) and the Conselho Nacional de Desenvolvimento Cient\'ifico e Tecnol\'ogico (CNPq) are acknowledged for the financial support.

\section*{Appendix A - Extending the ensemble}
In section \ref{representationloops}, we derived the effective field representation \eqref{effmixvortex} for an ensemble of uncorrelated loops. Here, we shall discuss how correlations between vortices are incorporated. Let us initially consider arrays with $V$ $N-$matching-points, with the probability amplitude \eqref{WF3}
\begin{align}
   \psi_{\{\gamma\}}  =  \xi_0^V\, \prod_{n=1}^I \psi_{\gamma_n}\;,
\end{align}
where $\psi_\gamma$ has the same form used in Eq. \eqref{psis}, and we also included a probability density $\xi_0^V$ for the ocurrence of the matching-points $\{ x_1, \dots , x_V \}$. Of course, the constraint $2I = N V$ must be satisfied, where $I$ is the number of lines. Now, 
from Eq. \eqref{Q1} and the approximation in Eq. \eqref{approxdiffusion}, the sum over lines with fixed initial point $x_1$ and final point $x_2$,  which carry a magnetic weight $\mathscr{C}$,    gives a factor  \begin{align}
     \int_0^\infty dL du_2 du_1 \, \tilde{\Psi}_0[\gamma(v,v_0,L)](E)\propto G_\mathscr{C}(x_2,x_1) \makebox[.5in]{,}
     O_\mathscr{C} \, G_\mathscr{C}(x_2,x_1)=\delta(x_2-x_1)\;.
     \label{greensrep}
 \end{align}
 In this respect, note that $Q$ contains the sum over all possible shapes with fixed length $L$, which is then supplemented by an integral over all possible $L$.   That is,  in the sum over  $\{\gamma\}$ within Eq. \eqref{we1}, the partial contribution of arrays with a given number of lines, fixed endpoints and topology, has the form
\begin{gather}
\propto \int d^3x_1 \dots d^3x_V \; \xi_0^{V}    \prod_{k=1}^I G_{\mathscr{C}_k } (x_2^k , x_1^k)  \;,
 \label{terms0}
\end{gather}
where the points $x_1^k$, $x_2^k$ ($k=1, \dots, I$) take values on the 
the set of vertices $\{ x_1, \dots , x_V \}$. 
It is clear that this  leads to the
wave functional 
\begin{align}
    &\tilde{\Psi}(E)=\prod\limits_{j=1}^N\int[D\bar{\phi}_j][D\phi_j] \exp \left[  -\int d^3x\, \left( \sum_{i=1}^N \bar{\phi}_i\, O_{\mathscr{C}_i} \phi_i  -\xi_0(\phi_{1}\dots \phi_{N} + {\rm c.c.}) \right)  \right]  
    \label{eqc}
  \;.
\end{align}
In effect,  Eq. \eqref{eqc} can be rewritten as
\begin{align}
    & \tilde{\Psi}(E)= \prod\limits_{j=1}^N (\det O_{\mathscr{C}_j})^{-1}  \exp  \int d^3x\, \xi_0  \left( \frac{\delta~}{\delta J_1} \dots \frac{\delta~}{\delta J_N} + \frac{\delta~}{\delta \bar{J}_1} \dots \frac{\delta~}{\delta \bar{J}_N} \right) \nonumber \\
   & \times \left. \exp \int d^3x\, d^3y\, \left(\sum_{i=1}^N  \bar{J}_i(x) G_{\mathscr{C}_i} (x,y) J_i(y) \right)
    \right|_{\bar{J}= J=0}  \;.
\end{align}
While the functional determinants give the center-vortex loop contribution $\tilde{\Psi}_0(E)$ in Eq. \eqref{effmixvortex}, 
the perturbative expansion (in $\xi_0$)
of the second factor gives rise to a superposition of terms of the form  \eqref{terms0}. Furthermore, 
being an (effective) field theory, this wave functional  
automatically fulfils  all the above mentioned requirements.
To include the contribution of monopoles, we introduce a parameter $\vartheta_0$ to describe the probability of their occurrence. That is, the sum over $\{\gamma\}$ in Eq. \eqref{we1} contains the partial contributions 
\begin{gather}
\propto \int d^3x_1 \dots d^3x_V \int d^3\bar{x}_1 \dots d^3\bar{x}_Z \; \xi_0^{V}   \vartheta_0^{Z}  \prod_{k=1}^I G_{\mathscr{C}_k } (x_2^k , x_1^k)  \;,
 \label{terms}
\end{gather}
for an array with $V$ points with $N$-line matching and $Z$ monopoles. Here, the corresponding sets of locations were denoted as $\{ x_1, \dots , x_V \}$ and $\{\bar{x}_1, \dots \bar{x}_Z\}$,  while the points $x_1^k$, $x_2^k$ ($k=1, \dots, I$) take values on them. Of course, the constraint $2I = N V + 2 Z$ must be satisfied. This leads to
\begin{align}
    &\tilde{\Psi}(E)=\prod\limits_{j=1}^N\int[D\bar{\phi}_j][D\phi_j] \exp \left[ -W(\Lambda)\right]\makebox[.4in]{,} W(\Lambda) = \int d^3x\left(-\frac{1}{3\kappa} \sum_{i=1}^N\bar{\phi}_i D^2(\Lambda^{\rm T}_{\mathscr{C}_i})\phi_i+V(\phi,\bar{\phi})\right)\;,\nonumber\\&
     V(\phi, \bar{\phi}) =\frac{\lambda_0}{2}\sum_i \left(\bar{\phi}_i \phi_i +\frac{\mu}{\lambda_0}\right)^2  -\xi_0(\phi_{1}\dots \phi_{N} +{\rm c.c.})-\vartheta_0\sum\limits_{i\neq j}\bar{\phi}_i\phi_{j} 
    \label{wvfunctional}\;.
\end{align}
\section*{Appendix B - Saddle point for a general eigenvalue $\Omega$}
\label{geneigen}
In section \ref{sectionwilson}, we obtained an approximate expression for the Wilson loop in a $k$-Antisymmetric representation in terms of a classical solution (Eq. \eqref{eqsaddle}), and computed it for the defining representation. In this section we shall study the saddle-point solution for a general $k-$Antisymmetric representation.
Let us begin by studying their properties. For $k=1$, it corresponds to the defining representation, which is spanned by the basis vectors $|\omega_1\rangle , \dots |\omega_N\rangle $. Their components are $|\omega_i\rangle = (0,\dots,1,\dots,0)^T$, with the nonzero entry being in the $i-$th position. The Cartan generators are diagonal in this basis, with eigenvalues given by the weights $\vec{\omega}_i$. For $k=2$, the representation is spanned by the  antisymmetrized tensor products
\begin{align}
    &|v_{ij}\rangle = \frac{1}{\sqrt{2}}\left(|\omega_i\rangle\otimes  |\omega_j\rangle-|\omega_j\rangle\otimes|\omega_i\rangle \right)\;,& i<j\;.
\end{align}
In this case, the generators and the weights are respectively given by
\begin{align}
    &\tilde{T}_A=T_A\otimes I + I\otimes T_A\;, \\&
    \vec{\Omega}_{(i_1,i_2)}=\vec{\omega}_{i_1}+\vec{\omega}_{i_2}\makebox[.5in]{,} i_1<i_2\;,
\end{align}
with $1 \leq i_1,i_2\leq N$. These results may be extended straightforwardly for $2<k<N$. For general $1\leq k < N$, the weights will then be
\begin{align}
    \vec{\Omega}_{(i_1,\dots,i_k)}=\vec{\omega}_{i_1}+\dots+\vec{\omega}_{i_k}\;,
\end{align}
 where $(i_1,\dots,i_k)$ is a tuple of integers satisfying $i_1<\dots<i_k$, $1\leq i_1,\dots,i_k \leq N$. The number of such weights is  $N(N-1)\dots(N-k)/k!$, which coincides with the dimension of the Antisymmetric representation with $N$-ality $k$. In order to identify the highest weight, a notion of ordering is necessary. As usual, we define a weight to be positive if the last nonvanishing component is positive. The weights of the defining representation satisfy
\begin{align}
    \vec{\omega}_1>\vec{\omega}_2>\dots>\vec{\omega}_N\;.
\end{align}
Based on these definitions, we can review the solution obtained in Ref. \cite{O-S-D-3d} for the highest weight $\vec{\Omega} \equiv \vec{\Omega}_{(1,\dots,k)}$ of the $k$-Antisymmetric representation. In this case, the matrix structure of the external source is given by
\begin{align}    2N\Omega|_{ij}=\langle \omega_i | 2N\Omega |\omega_j\rangle = \delta_{i(j)}2N\vec{\Omega}\cdot\vec{\omega}_{(j)}\;,
\end{align}
with no sum over $j$. This can be written as
\begin{align}
    {\rm diag}(2N\vec{\Omega}\cdot \vec{\omega}_1,\dots,2N\vec{\Omega}\cdot \vec{\omega}_N)=\frac{N-k}{N}P_1-\frac{k}{N}P_2\;, \label{projectors}
\end{align}
where $P_1={\rm diag}(1,1,\dots,0)$ with the $k$ first entries being 1 and the remaining being zero, while $P_2=I_N-P_1$. Here, we used that $\vec{\Omega}=\vec{w}_1+\dots+\vec{\omega}_k$ and the 
well-known relation \cite{conf-qg}
\begin{align}
    \vec{\omega}_{i}\cdot\vec{\omega}_{j}=\frac{N\delta_{ij}-1}{2N^2} \;.\label{scalarproduct}
\end{align} As the algebra of the matrices $P_1, P_2$ is closed, an Ansatz based on them closes the equations of motion. In particular, using
\begin{align}
    \Phi = (h_1P_1+h_2P_2)\,e^{i \theta_1\frac{N-k}{N}P_1-i\theta_2\frac{k}{N}P_2}
\end{align}
in Eq. \eqref{ww12}, we obtain scalar equations for the profiles $h_1,h_2,\theta_1,\theta_2$.
An alternative, equivalent form of this Ansatz is
\begin{align}
    \Phi=(\eta I_N+\eta_02N\Omega)\, e^{i2N\theta \Omega}e^{i\alpha}\;.
\end{align}
The equivalence is established by using the relations
\begin{align}
    \eta_0=h_1-h_2\makebox[.3in]{,}\eta=\frac{k}{N}h_1+\frac{N-k}{N}h_2\;.
\end{align}
To implement the boundary conditions
\begin{align}
    \Phi(x,y,z\to\infty)=v e^{i2N2\pi\Omega}I_N \makebox[.3in]{,}\Phi(x,y,z\to -\infty)=v I_N\;,
\end{align}
the profiles $\eta, \eta_0$ should satisfy
\begin{align}
    &\eta(-\infty)=\eta(\infty)=v\;, & \eta_0(-\infty)=\eta_0(\infty)=0\;.
\end{align}

As analyzed in Ref. \cite{O-S-D-3d}, in the relevant region of parameter space ($\lambda a^2, \xi v^{N-2} >> \vartheta$) the profiles $\eta, \eta_0, \alpha$ remain essentially constant at their vacuum values, and the transition between the different vacua is accomplished by a variation of $\theta$, which satisfies the equation
\begin{align}
    \partial^2_z \theta=\frac{\vartheta}{2}\sin \theta\;.
\end{align}
The energy per unit length of the soliton is then given by Eq. \eqref{ww11} and yields, after using Derrick's theorem,
\begin{align}
    \sigma_k=2v^2\, \frac{k(N-k)}{N}\int dz\,(\partial_z\theta)^2\;.
\end{align}
Finally, we study the solution for a general eigenvalue $\vec{\Omega}_{(i_1,\dots,i_k)}$ of the generators $D(T_q)$. In this case, the matrix $2N \Omega_{(i_1,\dots,i_k)}$ is given by
\begin{align}
     {\rm diag}( 2N \vec{\Omega}_{(i_1,\dots,i_k)}\cdot \vec{\omega}_1,\dots,2N \vec{\Omega}_{(i_1,\dots,i_k)}\cdot \vec{\omega}_N)=\frac{N-k}{N}P_{1(i_1,\dots,i_k)}-\frac{k}{N}P_{2(i_1,\dots,i_k)}\;.
\end{align}
The matrix $P_{1(i_1,\dots,i_k)}$ has zeros everywhere except on the diagonal entries $i$ which coincide with some of the $(i_1,\dots,i_k)$. Moreover, $P_{2(i_1,\dots,i_k)}=I_N-P_{1(i_1,\dots,i_k)}$. As the algebraic properties of these matrices (namely, their products and traces) are identical to those of Eq. \eqref{projectors}, the scalar equations obtained for the profiles are the same, and so is the expression for the energy. Therefore Eq. \eqref{resultk} holds for the general case.


\begin{thebibliography}{999}

\bibitem{indirectcenter}
L. Del Debbio, M. Faber, J. Greensite, Š. Olejník, Phys. Rev. D55 (1997) 2298.

\bibitem{ELRT} D. Engelhardt, K-. Langfeld, H. Reinhardt and O. Tennert, Phys.Lett. B431 (1998) 141.

\bibitem{langfeld_T}
M. Engelhardt, K. Langfeld, H. Reinhardt, O. Tennert, Phys. Rev. D61 (2000) 054504.
\bibitem{Reinhardt_topology}
H. Reinhardt, Nucl. Phys.  B628 (2002) 133.
\bibitem{reinhardtvortices}
M. Engelhardt, H. Reinhardt, Nucl.Phys. B567 (2000) 249.

\bibitem{nality}
 S. Kratochvila, P. de Forcrand, Nucl. Phys. B671 (2003) 103.
 \bibitem{randomsu2}
M. Engelhardt, H. Reinhardt, Nucl. Phys. B585  (2000) 591.


\bibitem{randomsu3}
M. Engelhardt, M. Quandt, H. Reinhardt, Nucl. Phys. B685 (2004) 227.

\bibitem{diff1}
L. E. Oxman, Phys. Rev. D98 (2018) 036018.

\bibitem{universe7080253}
D. R. Junior, L. E. Oxman, G. M. Sim\~oes, Universe
 7 (2021) 253.
        
\bibitem{Cosmai-2017} P.Cea, L. Cosmai, F. Cuteri, A. Papa,  Phys. Rev. D95 (2017) 114511. 

\bibitem{Yanagihara2019210} R. Yanagihara, T. Iritani, M. Kitazawa, M. Asakawa, T. Hatsuda,   Phys. Lett. B789 (2019) 210.

\bibitem{Kitazawa} R. Yanagihara, M Kitazawa, Prog. Theor. Exp. Phys. 9 (2019) 093B02.

\bibitem{su3} R. Yanagihara, M Kitazawa,   Prog. Theor. Exp. Phys. 7 (2020) 079201.

\bibitem{L_scher_2002}
M. Lüscher, P. Weisz, JHEP 7 (2002) 49. 

\bibitem{Rey} S. J. Rey, Phys. Rev. D40 (1989) 3396. 

 \bibitem{teper}
 B. Lucini and M. Teper, Phys. Rev. D 64 (2001) 105019.

\bibitem{4dlaw} B. Lucini, M. Teper, U. Wenger, JHEP 6 (2004) 012.



\bibitem{O-V} L. E. Oxman, D. Vercauteren,  Phys. Rev. D95 (2017) 025001.  
\bibitem{O-S} L. E. Oxman, G. M.  Sim\~oes, Phys. Rev. D99 (2019) 016011. 

\bibitem{O-S-J} D. R. Junior, L. E. Oxman, G.M.   Sim\~oes,
Phys. Rev. D102 (2020) 074005.


\bibitem{ambjorn} B. Durhuus, J. Ambj\o rn, T. Jonsson, \textit {Quantum Geometry: A Statistical Field Theory Approach}; Cambridge University Press: Cambridge, UK, 1997.

\bibitem{wheater} J. F. Wheater, J. Phys. A27 (1994) 3323. 


\bibitem{kleinert}
H. Kleinert, \textit{Path Integrals in Quantum Mechanics, Statics, Polymer
Physics, and Financial Markets}, World Scientific: Singapore, 2006.

\bibitem{diff3}
D.C. Morse, G.H. Fredrickson, Phys. Rev. Lett. 73 (1994) 3235.

\bibitem{fred}
Fredrickson, G.H. \textit{The Equilibrium Theory of Inhomogeneous Polymers}, 1st ed.; Clarendon Press: Oxford, UK, 2006; p. 452.


\bibitem{diff2}
L. E. Oxman, G. C. Santos-Rosa, B. F. I. Teixeira, Jour. Phys. A47 (2014) 305401.

\bibitem{oxmanreinhardt}
L. E. Oxman, H. Reinhardt, Eur. Phys. J. D78 (2018) 177.

\bibitem{O-S-D-3d}   D. R. Junior, L. E. Oxman, and G. M. Sim\~oes, JHEP 
01 (2020) 180.
\bibitem{directcenter}
    L. Del Debbio, M. Faber, J. Giedt, J. Greensite, Š. Olejník, Phys. Rev. D 58 (1998) 094501.

\bibitem{chains}
J. Ambjorn, J. Giedt, J. Greensite, JHEP 02 (2000) 033.
\bibitem{degrand}
T. A. DeGrand, D. Toussaint, Phys. Rev. D22 (1980) 2478.

\bibitem{conf-qg} L. E. Oxman, JHEP 03 (2013) 038.
\bibitem{refcho}
Y. M. Cho, Phys. Rev. D21 (1980) 1080; Phys. Rev. Lett. 46 (1981)
302; Phys. Rev. D23 (1981) 2415.
\bibitem{gen1}
L. E. Oxman, Phys. Rev. D82 (2010) 105020.

\bibitem{monopoles1}
H. Reinhardt, Nucl. Phys. B503 (1997) 505.

\bibitem{singer1}
D. Fiorentini, D. R. Junior, L. E. Oxman, R. F. Sobreiro, Phys. Rev. D105 (2022) 125015.
\bibitem{singer2}
D. Fiorentini, D. R. Junior, L. E. Oxman, G. M. Sim\~oes, R. F. Sobreiro, Phys. Rev. D 103 (2021) 114010.
\bibitem{topologicallattice}
Ph. de Forcrand, M. D’Elia, Phys. Rev. Lett. 82 (1999)
4582.
\end{thebibliography}
\end{document}